\documentclass[useAMS,usenatbib]{mnras}

\usepackage{newtxtext,newtxmath}
\usepackage[T1]{fontenc}
\usepackage{ae,aecompl}
\usepackage{amsmath}
\usepackage{float}

\usepackage{color}

\usepackage{graphicx}
\usepackage{dcolumn}
\usepackage{bm}
\usepackage{subcaption}
\usepackage{caption}
\usepackage{color}
\usepackage{hyperref}
\usepackage{makecell}
\usepackage{tabularx}
\usepackage{aas_macros}

\newcommand{\pd}{\partial}

\newcommand{\msun}{\,\mathrm{M}_\odot}
\usepackage[normalem]{ulem}

\title[ 2D Simulations of Fallback Supernovae]{Long-Time 2D Simulations of Fallback Supernovae: A Systematic Investigation of Explosions Dynamics and Mass Ejections}

\author[Sykes \& M\"uller]{
Bailey Sykes$^{1}$\thanks{E-mail: bailey.sykes@monash.edu}
and
Bernhard M\"uller$^{1}$\thanks{E-mail: bernhard.mueller@monash.edu}
\\
$^{1}$School of Physics and Astronomy, Monash University, VIC 3800, Australia
}

\begin{document}

\label{firstpage}
\pagerange{\pageref{firstpage}--\pageref{lastpage}}
\maketitle

\date{\today}

\begin{abstract}
We present a set of eight fallback simulations of zero-metallicity progenitors with masses between $60  \msun$ and $95 \msun$. The simulations are computed in 2D with the general relativistic  \textsc{CoCoNuT-FMT} code for the first few seconds after black hole formation, and then mapped  to the Newtonian code \textsc{Prometheus} for long-duration simulations beyond shock breakout. All simulations produce successful explosions with final energies ranging from $0.41\times 10^{51}\,\mathrm{erg}$ to $2.5 \times 10^{51}\,\mathrm{erg}$ and black hole masses from $20.7 \msun$ to  $34.4 \msun$. Explosion energies and remnant masses do not vary monotonically with progenitor mass, but the mass cuts cluster near the outer edge of the helium core. A supplementary model with decreased neutrino heating provides a tentative indication that successful explosions require the shock to reach the sonic point in the infall profile by the time of black hole formation. The propagation of the shock to the surface is only approximately captured by proposed shock invariants, but these may still be sufficient to extrapolate the final black hole mass from the first seconds of evolution. We also discuss potential multi-messenger signatures of the predicted fallback explosions. The enrichment of the ejecta in intermediate mass and iron-group elements varies considerably and is non-neligible for the more powerful explosions. Low-level neutrino emission after black hole formation from these very massive progenitors may be detectable in the case of a Galactic event.
\end{abstract}


\maketitle

\section{Introduction}

The fate of a dying massive star is difficult to predict. Such stars eventually undergo iron core collapse, leaving behind either a neutron star or black hole; however, it is not obvious which will occur for a given star and whether the collapse is accompanied by an explosion. Some approximate bifurcation between neutron-star forming and black hole forming supernovae can be made based on progenitor mass and composition \citep{Heger_Fryer_Woosley_Langer_Hartmann:2003}, but this is far from perfect \citep{Burrows_et_al:2024b} and determining the exact fate of stars with diverse structures and environments is still a challenge.

Based on the identified progenitors of Type~IIP supernovae,
there appears to be a paucity of explosion from stars with
zero-age main sequence (ZAMS) masses higher than $M_\mathrm{ZAMS} \gtrsim 18\msun$ in the local Universe \citep{Smartt:2015}, far lower than what
would be expected if all massive stars were to explode
as a CCSN. The likely implication of these observation is that such high-mass stars do not explode and instead quietly form black holes without luminous outburst; this is often referred to as a `failed' supernova. This population of failed CCSNe present a challenge to electromagnetic observations, although
one candidate for the disapperance of a massive star has now been identified \citep{adams_17,basinger_21}.
Failed supernovae may also become accessible to neutrino observatories and, in future, to gravitational wave detectors.

Studies of the collapse of massive stars have suggested that there are ``islands of explodability'' throughout the black hole forming range. However, there is currently no perfect predictor for successful explosions in this regime. Several studies have investigated the compactness parameter \citep{oconnor_ott:2011} as an indicator \citep[e.g.,][]{Ugliano_et_al:2012, Sukhbold_Woosley:2014,Sukhbold_et_al:2016,Mueller_et_al:2016}, but the critical value for an explosion to occur is still uncertain, and some studies have questioned whether compactness is a good predictor for explodability \citep[e.g.,][]{Burrows_et_al:2019}. Other metrics have been investigated, e.g., by \citet{Tsang_Vartanyan_Burrows:2022} in a machine learning framework,
but none of the proposed methods have proved sufficiently predictive yet when compared to multi-dimensional simulations.

In recent years, it has been increasingly realised that the distinction between successful supernova explosions and black hole forming collapse may not be a sharp one.  Around the dividing line, there may be \emph{fallback supernovae}, in which the shock is revived, but ongoing accretion onto the young neutron star triggers black hole formation at some stage of the developing explosion. After black hole formation, the stellar envelope may still be shed, or the shock may eventually turn around and fail to reach the stellar surface.
This fallback scenario is of significant interest observationally. Fallback supernovae may be responsible for peculiar abundance patterns in ultra metal-poor stars
\citep{nomoto_06,bessell_2015} or in X-ray binaries \citep{podsiadlowski_02}, which are indicative of pollution by a CCSN, but
with a deficit in iron-group elements. Moreover, there is evidence for black hole birth kicks \citep{Repetto:2012,Atri:2019}, which are most naturally explained by formation in a fallback supernova
with asymmetric mass ejection \citep{janka_13,Chan_Mueller_Heger_Pakmor_Springel:2018,Chan_Mueller_Heger:2020,mandel_20}. The black hole mass distribution also points to the existence of a fallback channel.
Gravitational-wave observations of very asymmetric mergers such as GW190814 and GW200210 \citep{LIGO_VIRGO:2019,abbott_23} have shown that there is a population of rather light black holes below $5 \msun$ and refuted earlier ideas \citep{oezel_12} of a mass gap between neutron stars and black holes. Such low black holes masses and the very high mass ratios in these
mergers are most readily explained by fallback supernovae \citep{antoniadis_22}.  There is also considerable interest in the high
end of the stellar-mass black hole distribution,
where stellar evolution had long predicted a mass gap due
to (pulsational) pair instability, which is not present in the observed distribution of binary black hole mergers \citep{LIGO_VIRGO:2020}.
Finally, it has long been speculated that some unusual observed supernovae may be fallback events. This includes faint events with very low nickel mass \citep{zampieri_03,moriya_10},
although most of these appear to be associated with lower progenitor mass, and may not be black hole formation events after all \citep{spiro_14}. Somewhat counterintuitively, fallback supernovae may instead be very bright events due to late-time powering of the light curve by accretion onto the black hole. A fallback origin has been proposed for supernovae of very high luminosity and/or very high ejecta mass such as OGLE-2014-SN-073 
\citep{moriya_18a,moriya_18b,moriya_19}. A somewhat related scenario, the unbinding of material from the stellar envelope due to the loss of gravitational mass by neutrino in black hole forming collapse, has also been discussed in the context of unusual transients as a possible explanation for luminous red novae \citep{Lovegrove_Woosley:2013, Byrne_Fraser:2022}.

Advances in CCSN modelling have already permitted first studies of the fallback scenario by means of multi-dimensional first-principle simulations. However, the number and scope of these studies still remain limited.
Several simulations have found shock expansion prior
to black hole formation in supernova models of very massive
progenitor stars in two (2D) and three dimensions (3D)
\citep{Chan_Mueller_Heger_Pakmor_Springel:2018,Chan_Mueller_Heger:2020,kuroda_18,pan_21,Rahman_Janka_Stockinger_Woosley:2022,Kuroda_Shibata:2023,Burrows_et_al:2024}, indicating
the possibility of explosions even in stars of very high compactness and with high binding energies. Only some of these simulations extend substantially beyond black hole formation, which is critical for determining the eventual fate of the incipient explosion, the birth properties of the black hole, and the observable transient resulting from the fallback supernova. Most notably, \citet{Chan_Mueller_Heger_Pakmor_Springel:2018}, 
\citet{Chan_Mueller_Heger:2020} (in 3D) and \citet{Rahman_Janka_Stockinger_Woosley:2022} (in 2D) have extended their simulations beyond shock breakout to quantify mass ejection, explosion energies, and black hole birth properties.

Aside from the long time scales in the problem, a major challenge in studying fallback supernovae lies in the importance of general relativistic effects. Black-hole formation is inherently a strong-field phenomenon and needs to be treated in general relativity at least for the purpose of validating the simulated dynamics around and shortly after collapse. While an approximate Newtonian treatment of gravity with a modified potential
\citep{Marek_et_al:2006,Mueller_Dimmelmeier_Mueller:2008}
is a well-tested approach in supernova simulations
for the proto-neutron star (PNS) phase \citep{Mueller_Janka_Dimmelmeier:2010}, a more rigorous approach is desirable for cases of black hole formation, but this significantly increases the complexity of the simulations, especially when a fully relativistic treatment is to be combined with multi-group neutrino transport.
This, however, bring the additional challenge of dealing with space-time metrics with singularities. The two principal methods for dealing
with black hole space-times, (moving) punctures
\citep{Brandt_Brugmann:1997,Campanelli:2006} and excision \citep{Thornburg:1993}, have been used extensively with free evolution schemes in the context of compact binary mergers.
In supernova simulations, such techniques are less common so far, but excision schemes have now been implemented in a number of supernova codes to study various aspects of black-hole formation \citep{Rahman_Janka_Stockinger_Woosley:2022,Sykes:2023,Kuroda_Shibata:2023}.

The existing studies in general relativity remain limited to
a small number of progenitors, however, and have yet to fully address key questions about the final fate of incipient supernova explosions after shock revival, and about the
explosion and remnant properties if the shock manages to
reach the surface. In this paper, we investigate some of these questions in greater depth using 2D simulations of eight different progenitors employing the new excision
capability of our \textsc{CoCoNuT-FMT} supernova code 
\citep{Sykes:2023}.
\begin{itemize}
    \item What physical criteria determine whether an incipient explosion survives after black-hole formation or whether the shock fails to reach the surface.
    \item If it explodes, how can the explosion and remnant properties be determined in the long term based on the initial state of the explosion around the time of black-hole formation?
    \item How do the explosion and remnant properties of black-hole forming supernovae depend on progenitor mass?
\end{itemize}
In addition, we investigate the neutrino emission and the composition of the ejecta as potential direct and indirect multi-messenger signals from fallback supernovae.
Given the recent interest in the origin of high-mass stellar black holes and transients with unusually high ejecta, we 
consider high-mass progenitors ($M_\mathrm{ZAMS} = 60 \texttt{-} 95\msun$) with zero metallicity (Population~III), which undergo little mass loss prior to collapse. As a side
effect, black-hole formation occurs rather early on in this
progenitors, which helps to make the simulations computationally feasible.
Although there are no such Population~III supernova
progenitors in the local universe, the models are sufficiently
representative of the high-mass tail of the stellar mass distribution in low-metallicity environments in the local universe. A significant population of such high-mass 
is already present at metallicities a few times below solar,
e.g., in the Large Magellanic Cloud \citep{Schneider_et_al:2018}.

This paper is structured as follows: In Section~\ref{sec:sim_overview}, we describe our simulation 
methodology and setup for relativistic simulations of
collapse, shock revival and black-hole formation
with  \textsc{CoCoNuT-FMT}, and for long-time
follow-up simulations up to shock breakout with the
Newtonian \textsc{Prometheus} code. We also describe
the progenitors models. In Section~\ref{sec:sim_results}, we perform a standard analysis of the CCSN explosion dynamics and the remnant properties, and the neutrino emission, and comment on the long-time evolution of the explosion until the emergence of the final mass cut. In 
Sections~\ref{sec:expl_criterion} and \ref{sec:analytic_shock},
we investigate the physical principles that determine whether the shock escapes the black hole and govern its propagation to the surface and the amount of mass ejection. We also discuss the ejecta composition in our models.
We summarise our results and present our conclusions in Section~\ref{sec:conclusion}.

\section{Simulation setup and numerical methods}
\label{sec:sim_overview}
\subsection{\textsc{CoCoNuT-FMT} simulations}
In this study, we simulate the collapse of eight massive stars with masses ranging from $60 \texttt{-} 95  \msun$. The simulations,
conducted with the CCSN simulation code \textsc{CoCoNuT-FMT}, begin at the onset of collapse and generally finish  several seconds after black hole formation before they are mapped to the
Newtonian \textsc{Prometheus} code for long-time evolution beyond shock breakout. Details of the \textsc{CoCoNuT-FMT} code and recent modifications are outlined in the following subsections.

\subsubsection{General relativity}

\textsc{CoCoNuT-FMT} uses an elliptic scheme for the Einstein equations based on the conformal flatness condition (CFC; \citealp{Isenberg:2008, Wilson_Mathews_Marronetti:1996}) in the modified  (xCFC) formulation of
\citet{Cordero-carrion_et_al:2009}. CFC and xCFC can be understood as approximate versions of the fully constrained formalism of the Einstein equations in the generalised Dirac gauge \citep{Bonazzola_Gourgoulhon_Grandclement_Novak:2004}.
The advantages of formulating the Einstein equations as a set of elliptic equations, as opposed to hyperbolic formulations such as in free-evolution schemes like the BSSN approach \citep{Baumgarte_Shapiro:1998}, consists in improved stability in long-term simulations in the presence of non-trivial matter fields and avoidance of stringent time-step limits for the dynamical evolution of the space-time metric.

As the (x)CFC approximation discards
gravitational wave degrees of freedom from the space-time, gravitational waves -- if required -- need to be extracted by means of the quadrupole formula
\citep{Finn_Evans:1990} with modifications for strong-field gravity \citep{Mueller_Janka_Marek:2013}. 

Following \citet{Sykes:2023}, and differing from standard formulations of the (x)CFC formulation \citep{CorderoCarrion_Vasset_Novak_Jaramillo:2014,Cordero-carrion_et_al:2009}, the elliptic equations for the conformal factor $\phi$,
the lapse function $\alpha$, and the shift vector $\beta^i$ are written as,
\begin{equation}
\label{eqn:elliptic_phi}
    \Delta \phi = -2 \pi \phi^{-1} \bigg[E^{*} + \frac{\phi^{6} K_{ij}K^{ij}}{16 \pi} \bigg],
\end{equation}
\begin{align}
\label{eqn:elliptic_alpha}
    \Delta \log(\alpha \phi) = & 2 \pi \phi^{-2} \bigg[E^{*} + 2S^{*} + \frac{7\phi^{6}K_{ij}K^{ij}}{16 \pi} \bigg] \nonumber \\
    &- \bigg( \frac{\partial \log(\alpha \phi)}{\partial r} \bigg)^{2},
\end{align}
\begin{equation}
\label{eqn:elliptic_beta}
    \Delta \beta^{i} + \frac{1}{3} \nabla^{i} \nabla_{j} \beta^{j} = 16 \pi \alpha \phi^{-2} (S^{*})^{i} + 2 \phi^{10} K^{ij} \nabla_{j} \frac{\alpha}{\phi^{6}},
\end{equation}
which ensures positivity of the lapse function $\alpha$ in the numerical solution scheme.
Here $K^{ij}$ is the extrinsic curvature, $E^{*}$ is the conformally rescaled energy density, $S^{*}$ is the trace of the stress tensor, and $(S^{*})^{i}$ is the momentum density. 

The CFC approximation uses maximal slicing in the bulk, i.e., by construction one has
\begin{equation}
\label{eqn:max_slicing}
    \gamma^{ij} K_{ij} = 0.
\end{equation}

The formation of black holes is dealt with using an excision scheme whereby the centre of the simulation grid is removed once the black hole forms (determined by the detection of an apparent horizon). The excised region is as large as possible while still being smaller than the apparent horizon to ensure no unphysical artefacts can propagate out of the excised region. 
We note that \citet{Kuroda_Shibata:2023} instead allow a ``buffer'' region between the apparent horizon and excision surface when defining their inner and outer regions.

Mathematically, the excision is handled by imposing appropriate boundary conditions on Equations~\eqref{eqn:elliptic_phi}-\eqref{eqn:elliptic_beta} once an apparent horizon develops. These are expressed as,
\begin{equation}
\label{eqn:bc_phi}
    \frac{\partial \phi}{\partial r} \bigg|_{r=R} = -\frac{M_{\textnormal{ADM}}}{2 R^{2}},
\end{equation}
\begin{equation}
\label{eqn:bc_alpha}
    \frac{\alpha}{\phi^{2}\beta^{r}} = \textnormal{const.},
\end{equation}
\begin{equation}
\label{eqn:bc_beta}
    \tilde{\gamma}^{ik} \nabla_{k} \beta^{j} + \tilde{\gamma}^{kj} \nabla_{k} \beta^{i} - \frac{2}{3} \tilde{\gamma}^{ij} \nabla_{k} \beta^{k} = 2 \alpha \phi^{-6} \hat{A}^{ij}.
\end{equation}
Here, $\hat{A}^{ij}$ is a conformally rescaled extrinsic curvature,
and the $M_{\textnormal{ADM}}$ term in Equation~\eqref{eqn:bc_phi} refers to the ADM mass of the star/black hole system, given by \citet{Gourgoulhon:2007} as,
\begin{equation}
\label{eqn:full_adm}
     M_{\textnormal{ADM}} = \frac{1}{16 \pi} \lim_{\mathcal{S} \rightarrow \infty} \oint_{\mathcal{S}} \left[\nabla^{j} \gamma_{ij} - \nabla_{i} (f^{kl} \gamma_{kl} )\right) s^{i} \sqrt{q} \,d^{2}y ,
\end{equation}
where $\mathcal{S}$ is a sphere (in our case the outer boundary of the grid) with induced surface element $\sqrt{q}\,d^{2}y$ and $s^{i}$ is an outward-pointing unit normal to $S$. 

Equations~\eqref{eqn:bc_alpha} and \eqref{eqn:bc_beta} apply at the excision boundary (which is in practice \textit{almost} the same as the apparent horizon), while Equation~\eqref{eqn:bc_phi} is a boundary condition at the \textit{outer} boundary of the computational grid and ensures conservation of the ADM mass. There is no angular dependence to these boundary conditions as the metric is currently computed in spherical symmetry.

Black hole growth is accounted for by dynamically updating the metric boundary conditions as the apparent horizon expands. We find this scheme to be stable in all test cases for several seconds after black hole formation; it is likely that even longer simulations could be performed with this scheme.

\subsubsection{Hydrodynamics}
The hydrodynamics solver in \textsc{CoCoNuT-FMT} has been documented extensively over the course of its development \citep{Dimmelmeier_Font_Mueller:2002, Mueller_Janka_Dimmelmeier:2010, Muller_Janka:2015}. Its key features are a high-resolution shock-capturing scheme with piecewise-parabolic reconstruction \citep{Colella_Woodward:1984} and a hybrid HLLC/HLLE \citep{Toro_Spruce_Speares:1994, Mignone_Bodo:2005, Harten_Lax_vanLeer:1983, Einfeldt:1988} Riemann solver to enable accurate solutions to the relativistic fluid equations in spherical polar coordinates. 

Following \citet{Sykes:2023}, the hydrodynamics module has been modified to solve the internal energy equation in highly supersonic flow around the black hole, and an appropriate extrapolation scheme is used at the excision surface to define outflow boundary conditions.

We also exploit the ability of the code to
compute the inner region of the grid in spherical symmetry. Spherical averaging is used in the first five grid cells outside the apparent horizon.
This has two main benefits: Firstly, it avoids non-spherical effects at the excision boundary which would not interact well with the metric boundary conditions because they assume spherical symmetry. Secondly, some extreme hydrodynamic conditions adjacent to the excision surface that can cause stability issues are smoothed out. The material being spherically averaged is accreting near the speed of light, and is well within the sonic point; consequently, there is effectively no impact of this procedure on the collapse dynamics.

\subsubsection{Neutrinos}
For the neutrino transport, we use the fast multi-group transport (FMT) scheme of \citet{Muller_Janka:2015}. 
The FMT scheme solves the stationary transport equation using the ray-by-ray approximation and disregards most velocity-dependent terms. 
Different from the original FMT scheme, which has an inner boundary condition of zero flux at the centre, we impose zero outgoing flux on the apparent horizon after black hole formation. 
The FMT scheme provides a good balance between accuracy and computational efficiency, which is a priority in the case of simulations of core collapse up to several seconds after black hole formations.

\subsubsection{Equation of state}
At high densities, we use the equation of state (EoS) of \citet{Lattimer_Swesty:1991} with a
bulk compressibility modulus of $K=220\, \mathrm{MeV}$
(LS220). Different equations of state have been shown to have a marked impact on the time of black hole formation post-bounce \citep[e.g.][]{oconnor_ott:2011,Powell_Mueller_Heger:2021}, with LS220 being conducive to somewhat earlier black hole formation than the current standard choice (SFHo, \citealp{Steiner_Hempel_Fischer:2013}) for most CCSN simulations. The LS220 EoS is actually marginally ruled out by the most recent set of observational and experimental constraints \citep{Oertel_Hempel_Klahn_Typel:2017, Fischer_et_al:2014, Lattimer:2019}, but the discrepancy is small. For the purpose of this project, somewhat earlier black hole formation with LS220 compared to, e.g., the SFHo EoS, is advantageous from a practical point of view and is not expected to affect the key point of interest, i.e., the physical conditions and mechanisms that determine the outcome of fallback supernovae with black hole formation.

Nuclear burning is accounted for using the flashing treatment of \citet{Rampp_Janka:2002}.

\subsection{PROMETHEUS simulations}

We also perform long-duration simulations to shock breakout to complement our \textsc{CoCoNuT-FMT} simulations. These not only allow us to confidently identify which models undergo mass ejection, but also determine the ejecta mass, composition, and the final explosion energy of the CCSNe. By calculating these quantities in a rigorous manner, we are then able to make comparisons to analytic theory for weak shock propagation and assess the performance of these tools as a predictive method.
In a similar manner to \citet{Chan_Mueller_Heger_Pakmor_Springel:2018, Chan_Mueller_Heger:2020} and \citet{Rahman_Janka_Stockinger_Woosley:2022}, we map the simulation grids of our models to a non-relativistic hydrodynamics code and replace the black hole with a sink particle and an inner boundary condition for free inflow
onto the sink particle at a radius considerably larger than the Schwarzschild radius.
We use the
\textsc{Prometheus} code \citep{Fryxell_Mueller_Arnett:1991, Mueller_Fryxell_Arnett:1991}, which supports a moving grid to expedite long-time simulations to bridge the temporal and spatial scales from black hole formation at the centre to shock breakout days or weeks later.

\subsubsection{Mapping to \textsc{Prometheus}}

Unlike \textsc{CoCoNuT-FMT}, \textsc{Prometheus} employs a Newtonian treatment of gravity and, as such, it is important that GR effects on the dynamics of the accretion flow are already minimal before switching codes. We therefore map several seconds after black hole formation when the transient effects of black hole collapse have passed and a quasi-stationary supersonic inflow has developed behind the ejecta. The mapping time is much longer than the free-fall timescales at the sonic radius, which also guarantees that the shock is not in sonic contact with the region where relativistic effects are important. The exact criterion for mapping is that the maximum radius of the (asymmetric) shock is at least $5 \times 10^{5} \, \mathrm{km}$; this should provide as long a simulation as possible without incurring interactions with the outer boundary which may artificially distort the shock.

The \textsc{Prometheus} grid has an inner and outer boundary initially set to $10^{3} \, \mathrm{km}$ and $2 \times 10^{9} \, \mathrm{km}$, respectively. The outer boundary is set such that it encompasses the entire progenitor star, and sits in a very low-density artificial isothermal atmosphere. As the shock approaches the outer boundary, the grid is expanded outwards as needed, and the new zones filled with the low-density atmosphere \citep{Mueller_et_al:2018}. 
The inner boundary sits at a radius safely within the sonic radius (specifically strictly below the point where the inflow Mach number $\mathcal{M}$ satisfies $\mathcal{M} > 2$), and imposes an outflow boundary condition\footnote{Physically, material is allowed to flow inwards over the boundary towards the excised black hole, but \textit{out} of the simulation grid, hence the term ``outflow''.}. The inner boundary is moved radially outwards as the shock expands, with a corresponding central black hole point mass increasing with the mass flux over the boundary. By moving the inner boundary, the minimum time step set by the Courant–Friedrichs–Lewy condition is allowed to increase, hence allowing simulation over timescales of several orders of magnitude.

Our mapping process involves three distinct processes depending on the radial coordinate:
\begin{itemize}
    \item $r < 10^{10} \, \mathrm{cm}$: The \textsc{CoCoNuT} grid is interpolated onto the new \textsc{Prometheus} grid.
    \item $10^{10} < r < R_{*}$: The progenitor data are smoothly interpolated onto the new grid up to the stellar radius, $R_{*}$.
    \item $r > R_{*}$: The remainder of the grid is filled with an exponentially decaying atmosphere of an artificial nuclear dummy species. The dummy species allows us to track the expansion of the stellar surface, and to identify and remove swept-up matter from the numerical atmosphere in our analysis and for potential radiative transfer follow-up simulations in the future.
\end{itemize}

\subsection{Progenitor models}
The chosen progenitors are listed in Table~\ref{tab:prog_models}. All models have been computed with  the stellar evolution code \textsc{KEPLER} \citep{KEPLER:1978} and have zero metallicity, representative of Population~III stars in the early Universe. The mass range covers the most massive progenitors that directly evolve to iron core-collapse, as well as the lower end of the pulsational pair-instability regime with iron core collapse after one or several pulses.

\newcolumntype{Y}{>{\centering\arraybackslash}X}
\begin{table}
\begin{center}
    \begin{tabularx}{\columnwidth}{Y|YYYYY}
    \hline
        Model & 
        \makecell{$M_{\mathrm{ZAMS}}$ \\ ($\msun$)} & 
        \makecell{$t_{\mathrm{bounce}}$ \\ (ms)} & 
        \makecell{$t_{\mathrm{BH}}$ \\ (ms)} & 
        \makecell{$t_{end}$ \\ (ms)} &
        $\xi_{2.5}$\\
    \hline
        z60 & 60 & 392 & 2008 & 4320 & 0.624 \\
        z65 & 65 & 402 & 2244 & 4257 & 0.627 \\
        z70 & 70 & 442 & 896 & 3497 & 0.708 \\
        z75 & 75 & 456 & 783 & 4191 & 0.766 \\
        z80 & 80 & 472 & 670 & 4517 & 0.827\\
        z85 & 85 & 524 & 327 & 6235 & 1.17\\
        z90 & 90 & 548 & 263 & 6591 & 1.33\\
        z95 & 95 & 470 & 319 & 6634 & 1.18\\
    \hline
    \end{tabularx}
\caption{Summary of the progenitor models used in this study. The mass is the zero-age main sequence mass.
$t_{\textnormal{bounce}}$ is the time of bounce relative to the onset of collapse. $t_{\textnormal{BH}}$ is the time of black hole formation relative to bounce, and $t_{\mathrm{end}}$ is the total simulated post-bounce time in \textsc{CoCoNuT-FMT}. The last column gives the value of the compactness parameter $\xi_{2.5}$ \citep{oconnor_ott:2011} at bounce.
}
\label{tab:prog_models}
\end{center}
\end{table}

A tentative prediction for the explodability of these models based on their compactness parameter \citep{oconnor_ott:2011}, $\xi_{2.5}$, indicates it is unlikely that any will explode. In all cases $\xi_{2.5} > 0.6$, which is significantly larger than the typical critical value of $0.2-0.3$ favoured by parameterised studies of massive star explodability (e.g. \citet{Mueller_et_al:2016, Sukhbold_et_al:2016}). We note that the zero-metallicity stars we consider here may be easier to explode due to differences in core structure and burning during the pre-collapse phase \citep{Sukhbold_Woosley:2014}. 

This range of $\xi_{2.5}$ is shifted upwards compared to those obtained by \citet{Aguilera-Dena_et_al:2023} -- although some differences are naturally expected between their variable-metallicity stripped envelope progenitors and our zero-metallicity Pop-III models. While the non-monotonicity of $\xi_{2.5}$ is a well-established property \citep{oconnor_ott:2011}, our results show an upward (almost monotonic) trend with increasing $M_\mathrm{ZAMS}$. This was also observed by \citet{Sukhbold_Woosley:2014} in their low-metallicity models. The non-monotonicity in $\xi_{2.5}$ as a function of ZAMS mass is mostly due to changes in shell structure and the transitions from convective to radiative burning of C and O as the core gets larger.
These transitions give rise to peaks in $\xi_{2.5}$ at ZAMS masses around $20 \msun$ and $30 \texttt{-}40 \msun$, respectively.
Our models are from a higher mass range where the shell structure of stars evolving towards core collapse becomes much more stable. Consequently, the compactness parameter tends to increase quite smoothly, at least until the ZAMS mass reaches the pair instability regime, where pair instability pulses can impact the stellar structure (as is particularly evident for the z90 model).

\section{Overview of black hole formation simulations}
\label{sec:sim_results}
In this section we provide an overview of the collapse and explosion dynamics for our \textsc{CoCoNuT-FMT} simulations, and also analyse the predicted neutrino emission before and after black hole formation.

\subsection{Shock trajectory}
\begin{figure*}
    \centering
    \includegraphics[width=\textwidth]{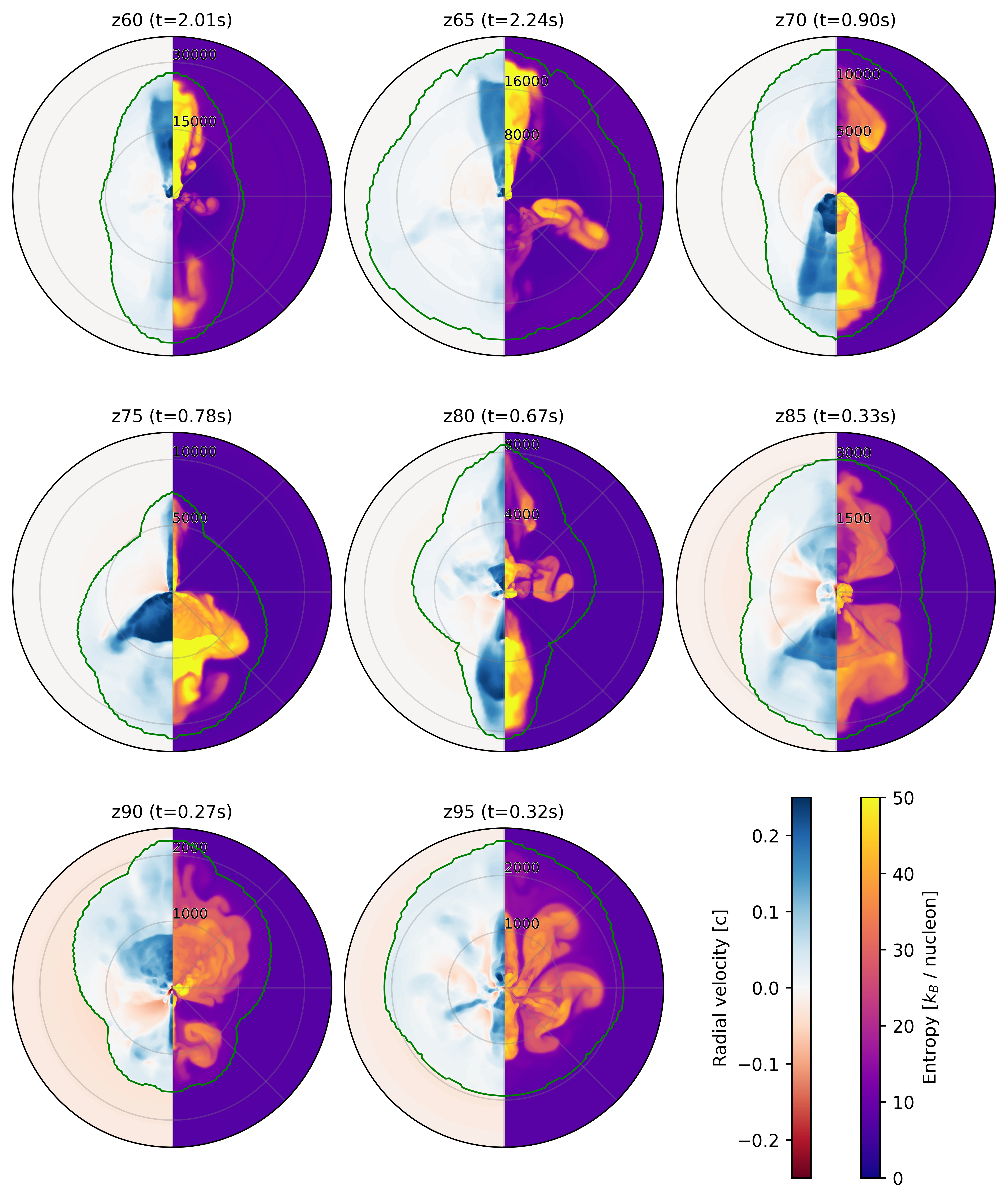}
    \caption{Snapshots of the simulations at their respective times of black hole formation. The left side of each plot is the radial velocity, with the corresponding colour bar on the left in the bottom right panel of the figure. The entropy is shown on the right, with the matching colour bar on the right in the bottom right panel. The spatial scale has been chosen to show the entire shock (which is denoted by a green line in each snapshot) and hence is different for each model; the radial coordinate can be read off from the markings along the vertical axis. The post-bounce time is shown at the top of each snapshot.}
    \label{fig:vex_sto_2d}
\end{figure*}

In our \textsc{CoCoNuT-FMT} simulations, we find shock revival for all models, followed by black hole formation between $0.12 \, \mathrm{s}$ to $2.1 \, \mathrm{s}$ later. The ejecta geometry, visualised by the specific entropy and radial velocity, and the shock geometry (green line) at the time of black hole formation are shown in Figure~\ref{fig:vex_sto_2d}. By this stage, most explosions have developed at least a moderately dipolar structure; this structure is aligned with the grid axis due to the implicit azimuthal symmetry in 2D. The post-bounce times at the time of black hole formation are also shown. In general, larger delays before black hole formation allow the shock to expand further. This produces a range of shock radii at the time of black hole formation.

Radial velocities overall show moderately fast expanding material, with pockets of infalling/accreting material. Very fast outward flows along the axis are driven by neutrino heating. These are accentuated by artificial squeezing out of accreted material due to the enforced symmetry, but the presence of the outflows per se is not an artefact of the geometry. Plumes of hot material with high entropy ($\sim 40 \, k_\mathrm{B} / \mathrm{nucleon}$) and moderate ($\sim 0.1c$) outwards velocities are seen in all models. One can clearly recognise that reverse shocks have formed along some directions already, perhaps most prominently in the lower hemisphere of the z75 model.

\begin{figure*}
    \includegraphics[width=\linewidth]{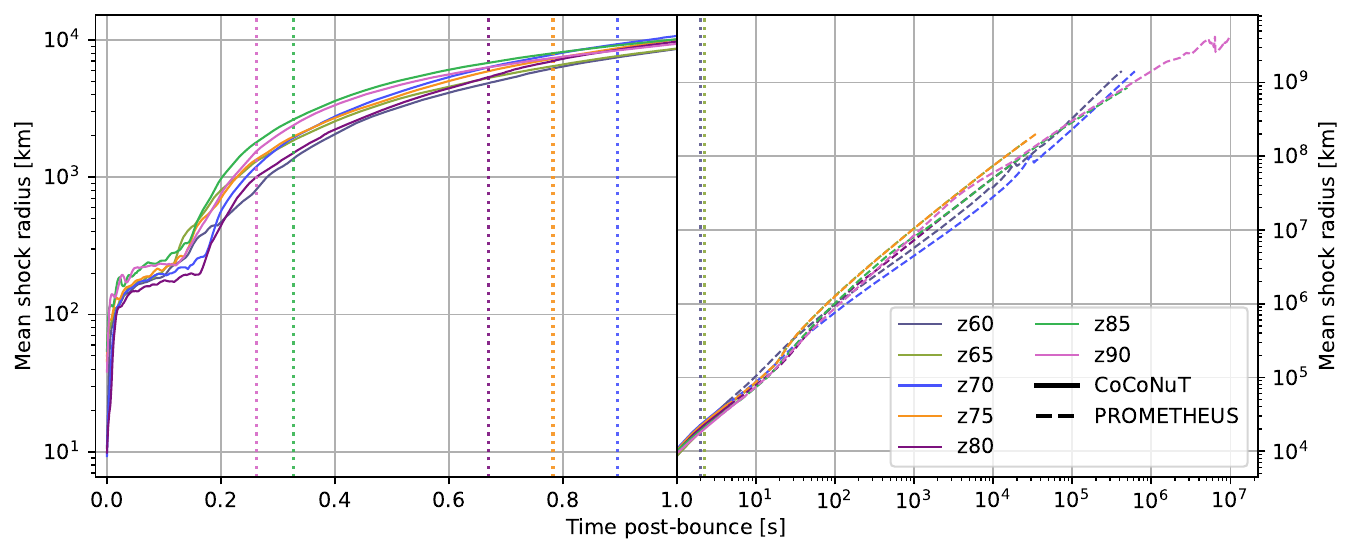}
    \caption{Trajectory of the shock after bounce. The mean radius of the shock is shown, although this can be skewed by the somewhat artificially advanced shock along the poles in some simulations. Dotted vertical lines denote the time of black hole formation for the model with the corresponding colour; the exact times are given in Table \ref{tab:prog_models}. Solid lines show the results from the \textsc{CoCoNuT-FMT} simulations, while dashed lines show the \textsc{Prometheus} results. The start and end times of the \textsc{Prometheus} simulations are not the same for each model, however the black hole has always formed $>1 \, \mathrm{s}$ before mapping to the Newtonian code.}
    \label{fig:coco_prom_shock_r}
\end{figure*}

Once the explosion phase begins, the shock propagates approximately linearly with time, and the trajectory (Figure~\ref{fig:coco_prom_shock_r}) exhibits a similar shape for all models. 

The shock trajectory also shows no signs of change in response to the collapse of the PNS to a black hole for any of the simulated models. While the formation of the black hole very quickly (on the order of milliseconds) establishes an infall profile inside the sonic point, the changes outside this radius are less dramatic. Any sign of collapse which manifests itself outside the sonic point of the infall region must then traverse the distance to the shock to have an impact. This distance is quite large: i.e., the log-ratio of mean shock radius and mean sonic point radius, $\log_{10}(r_{\mathrm{shock}} / r_\mathrm{sonic})$ is between $1.6$ and $2.7$ depending on the specific model. This makes the `instantaneous' sound crossing timescale as large as $10^{2} \texttt{-} 10^{3} \, \mathrm{ms}$. However, such an approximation neglects the continuing shock expansion, which is quite significant over these timescales, with Mach numbers of $0.3-0.7$ in the post-shock flow. Consequently, any signals of collapse take even longer to reach the shock, and may result in the shock ultimately being out of sonic contact with the black hole.

At the time of black hole formation, the direct impact of GR on the shock dynamics are small, with values of $GM/Rc^{2}$ less than $2 \times 10^{-3}$ at the shock radius in all models. The accretion region within about $10 \, \mathrm{km}$ of the black hole exhibits $GM/Rc^{2} \gtrsim 0.1$, consistently over the course of the simulation. At late times, the growth of the black hole pushes the relativistic accretion region outwards by a factor few in the most massive models, hence expanding the region to tens of kilometres.

The effect of relativistic accretion near the black hole on the explosion dynamics is tempered by this region being inside the sonic point of the infall region. This means that changes to the hydrodynamic properties of the infall region due to GR are unable to propagate through sound waves to the exterior region. For this reason, it is helpful to consider the magnitude of GR effects at the sonic point also. This tends to be in the range of $GM/Rc^{2} \sim 0.01-0.1$. While these values do not indicate a particularly strong GR effect, they are close enough to still suggest that inclusion of GR may be helpful in high-accuracy simulations.

Furthermore, GR effects play a major role in determining the state of the explosion at the time of black hole formation.
Since GR is known to impact the neutrino emissions of the PNS \citep{Mueller_Janka_Marek:2012}, GR can alter the neutrino heating in the gain region as compared to a Newtonian treatment, and hence drive the shock at different rates of expansion.

GR effects also remain important for the weak neutrino emission after black hole formation in the relativistic accretion region. This may be relevant both to the detection of neutrino transients from a Galactic supernovae (e.g., for identifying accretion properties of nascent black holes in a black hole forming Galactic supernova) and to the detection of the diffuse supernova neutrino background \citep{beacom_10,Scholberg:2012, Suliga:2022}.

\subsection{Explosion energy}
Following previous works  \citep{Mueller_Janka_Marek:2012, Buras_Rampp_Janka_Kifonidis:2006}, we define a diagnostic explosion energy to track the energy of formally unbound ejecta. 
If, as in our models, the unshocked outer shells have a significant binding energy (``overburden''), the diagnostic energy at early times is not a good predictor for the asymptotic explosion energy, but remains useful to quantify the dynamics of the explosion as it evolves.
To obtain asymptotic explosion energies, we map the 
\textsc{CoCoNuT-FMT} simulations to the Newtonian
\textsc{Prometheus} code after following the evolution
past black hole formation for a significant amount
of time in \textsc{CoCoNuT-FMT} to minimise artefacts
from the transition to a Newtonian treatment.

The diagnostic explosion energy takes the form,
\begin{equation}
    E_{\mathrm{expl}} = \int_{D} \rho W \mathrm{e_{bind}} \,\mathrm{d}\tilde{V},
    \label{eqn:diag_energy}
\end{equation}
where $d\tilde{V}$ is the curved-space volume element, $\rho$ is the baryonic rest-mass density and $W$ is the Lorentz factor
(which must be set to $W=1$ in the Newtonian approximation). The $\mathrm{e_{bind}}$ term in the Newtonian limit is the sum of internal, kinetic and potential energies (minus rest mass energy):
\begin{equation}
    e_{\mathrm{bind}} = \epsilon + \frac{v^{2}}{2} + \Phi.
\end{equation}
In GR, this takes on a more complicated form \citep{Mueller_Janka_Marek:2012},
\begin{equation}
    e_{\mathrm{bind}} = \alpha \bigg[\bigg(c^{2} + \epsilon + \frac{P}{\rho}\bigg)W^{2} - \frac{P}{\rho}\bigg] - Wc^{2}.
    \label{eqn:gr_bind}
\end{equation}
To avoid double-counting the potential of material outside a given radius, the Newtonian gravitational potential outside that radius is subtracted from $\mathrm{e_{bind}}$ \citep{Mueller_Melson_Heger_Janka:2017}. The integration domain, $D$, simply covers the cells with positive binding energy ($\mathrm{e_{bind}} > 0$), and $v_{r} > 0$ to avoid including any energetic accreting material. Close to the horizon, the space-time is non-stationary, and the factor $\alpha$ in Equation~(\ref{eqn:gr_bind}) no longer correctly accounts for the gravitational binding energy. Failing to filter out the accretion region can therefore give rise to spuriously high values of the explosion energy.

\begin{figure*}
    \centering
    \includegraphics[width=\linewidth]{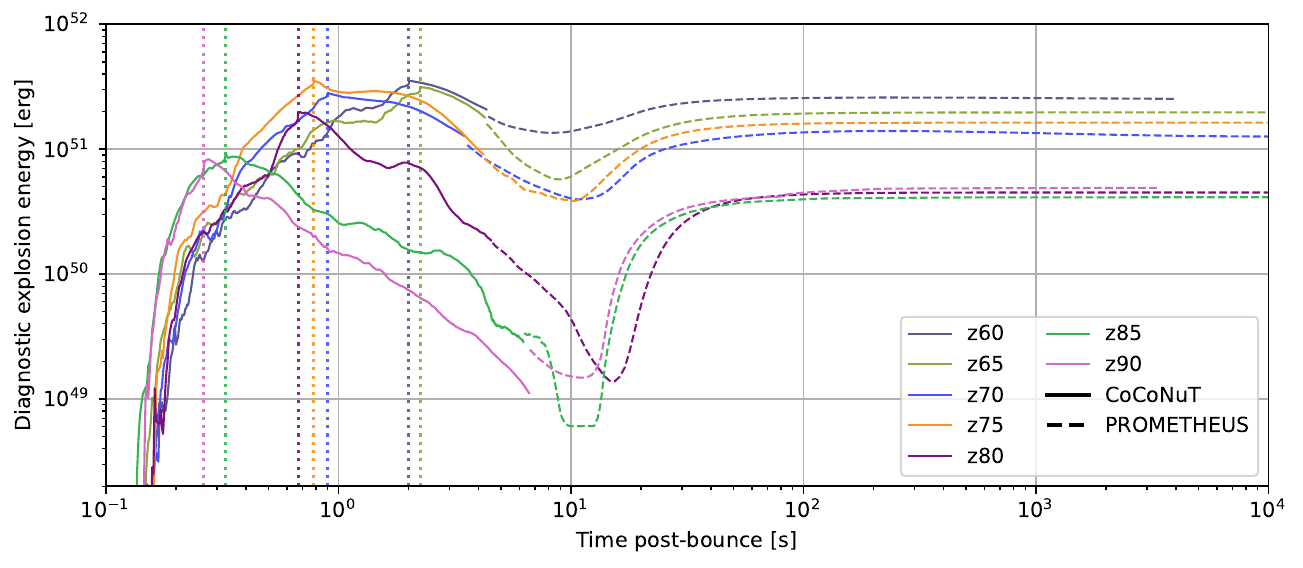}
    \caption{Diagnostic explosion energies for all models. The dotted vertical lines again denote the time of black hole formation for the respective model. Solid lines show the results for the \textsc{CoCoNuT} simulations, while the dashed lines show the explosion energy calculated from the Newtonian follow-up simulations with the \textsc{Prometheus} code.}
    \label{fig:exp_eng}
\end{figure*}

The explosion energies (Figure~\ref{fig:exp_eng}) show a sudden change in slope at the time of black hole formation (shown by vertical dashed lines), going from a positive to negative gradient of similar magnitude. Part of this feature can be explained by the immediate cut-off of neutrino heating once the black hole forms (and promptly accretes the neutrinosphere). Neutrino heating was previously the dominant mechanism driving up the explosion energy, and so when it stops,
further growth of the explosion energy ceases.
Furthermore, once the black hole forms, the explosion energy starts to decrease due to the accumulation of bound matter by the shock and
$P\,\mathrm{d}V$ work by the ejecta (though
$P\,\mathrm{d}V$ work by bound material can increase the explosion energy again later). This behaviour has already been seen in simulations by \citet{Chan_Mueller_Heger_Pakmor_Springel:2018} as well as \citet{Rahman_Janka_Stockinger_Woosley:2022}.

Diagnostic explosion energies do not match exactly between the \textsc{CoCoNuT} and \textsc{Prometheus} simulations at the time of grid mapping for a number of reasons. Firstly, the Newtonian explosion energy is lower than the GR equivalent in most models; this is a result of Newtonian and relativistic energies not being equivalent. As such, some marginally unbound material on the \textsc{CoCoNuT} grid (e.g. the fast, axial jet-like structures present in many models) may become tenuously bound after the mapping. Additionally, when moving to the Newtonian calculation, some material outside the \textsc{CoCoNuT} computational grid (which includes only the innermost $10^{5} \, \mathrm{km}$ of the progenitor) typically becomes spontaneously unbound when the \textsc{Prometheus} simulations start. This is due to shells in the progenitor structure (which may already be lightly bound, e.g., after
strong expansion following a shell burning pulse)
experiencing a sudden reduction in gravitational binding owing to the mass-energy lost to neutrinos during the GR simulation of the core. The decrease of the gravitational mass compared
to the pre-collapse stage unbinds these marginally bound shells and produces a net contribution to the diagnostic explosion energy \citep{Lovegrove_Woosley:2013}.

Another effect is particularly prominent in the z90 model, which differs from the others in that the diagnostic explosion energy increases after the mapping to \textsc{Prometheus}. 
This model has a relic pair-instability pulse in the progenitor at a radius of $6 \times 10^{7} \, \mathrm{km}$. While other models show similar features in various places, none are as strong as in the z90 model. The diagnostic explosion energy of this pulse alone (it is also the only unbound material in the progenitor) is $2.5 \times 10^{49} \, \mathrm{erg}$, which almost exactly matches the discrepancy between the \textsc{CoCoNuT-FMT} and \textsc{Prometheus} energies at the time of mapping.

These disagreements between the codes are expected, to a large extent unavoidable, and ultimately are not cause for much concern, as the long-term behaviour, i.e., the final explosion energy, is more important. Furthermore, the explosion energies are global quantities relying on the difference of large numbers: the energy of the material itself and the overburden. This naturally leads to some numerical inaccuracy in its calculation. However, the behaviour of the matter is dictated locally by the usual hydrodynamics equations, and it is not possible to determine if a given cell (in isolation) is formally unbound or not at any point. This means the disagreement between energies has no effect on the dynamics and is largely an artefact of the analysis amplified by the nature of the calculation.

All models reach a stable final explosion energy a few minutes after the onset of collapse. Less massive models tend to have higher explosion energies while more massive models tend to be less energetic, although the trend is not exact.
We do not quantify this trend, and do not attempt a comparison to observational data. Such a comparison is currently only possible for CCSNe from progenitors of lower mass \citep{Murphy_Mabanta_Dolence:2019}, where appropriate samples of observed transients are available \citep[e.g.,][]{Pejcha_Thompson:2015}. We note, however, that our results
for high-mass progenitors are indicative of a negative slope of the explosion energy with ZAMS mass rather than the positive correlation that has been suggested for progenitors of lower mass both by
observations \citep{Poznanski:2013,Murphy_Mabanta_Dolence:2019} and simulations \citep{Burrows_et_al:2024b, Mueller_et_al:2016}.

\citet{Rahman_Janka_Stockinger_Woosley:2022} explore a similar region of the ZAMS parameter space, but find very weak explosion energies, and consequently make no conclusions on the final progenitor-explosion energy connection.

The final explosion energies range from $4.1 \times 10^{50} \, \mathrm{erg}$ to $2.5 \times 10^{51} \, \mathrm{erg}$. At the low end, these are fairly typical CCSNe explosion energies, while those at the higher end are relatively energetic compared to the general population \citep{Murphy_Mabanta_Dolence:2019}. CCSNe with explosion energies as high as, say, the z60 model, are not unheard of however; for instance \citet{Terreran_et_al:2017} find OGLE-2014-SN-073 has an explosion energy in excess of $10^{52} \, \mathrm{erg}$, larger than any of our models. While they conclude that such large energies cannot be explained by the usual neutrino mechanism, the extended hydrogen envelope they infer is similar to those of our Pop-III model. This suggests that large explosion energies may be accessible to this class of progenitor star, if additional effects such as magnetic field and realistic progenitor rotation were included.

\newcolumntype{Y}{>{\centering\arraybackslash}X}
\begin{table}
\begin{center}
    \begin{tabularx}{\columnwidth}{Y|YYYYY}
    \hline
     &  & \multicolumn{2}{c}{$M_\mathrm{PNS}$} & \multicolumn{2}{c}{$M_\mathrm{BH}$} \\ 
    Model & $E_{\mathrm{BH}}$ & Baryonic & Grav. & Baryonic & Grav. \\
    \hline
    z60 & 3.5 & 2.33 & 2.14 & 2.76 & 2.39 \\
    z65 & 3.1 & 2.33 & 2.13 & 2.74 & 2.36 \\
    z70 & 2.8 & 2.38 & 2.24 & 3.51 & 3.21 \\
    z75 & 3.5 & 2.39 & 2.26 & 3.78 & 3.50 \\
    z80 & 2.0 & 2.42 & 2.30 & 3.60 & 3.33 \\
    z85 & 0.92 & 2.51 & 2.41 & 8.51 & 8.34 \\
    z90 & 0.80 & 2.55 & 2.47 & 9.65 & 9.49 \\
    \end{tabularx}
    \caption{Select results of the \textsc{CoCoNuT-FMT} simulations. $E_{\mathrm{BH}}$ is the diagnostic explosion energy at the time of black hole formation. Final PNS masses are listed under $M_{\mathrm{PNS}}$, both as baryonic masses, and approximate gravitational (grav.) masses. black hole masses at the end of the respective simulations are also listed, firstly as baryonic masses calculated by mass conservation on the grid, and then again as an approximate gravitational mass. Energies are given in units of $10^{51} \, \mathrm{erg}$ and masses in solar masses ($\msun$).}
    \label{tab:coconut_results_table}
\end{center}
\end{table}

\subsection{Remnant properties}
\begin{figure}
    \centering
    \includegraphics[width=\linewidth]{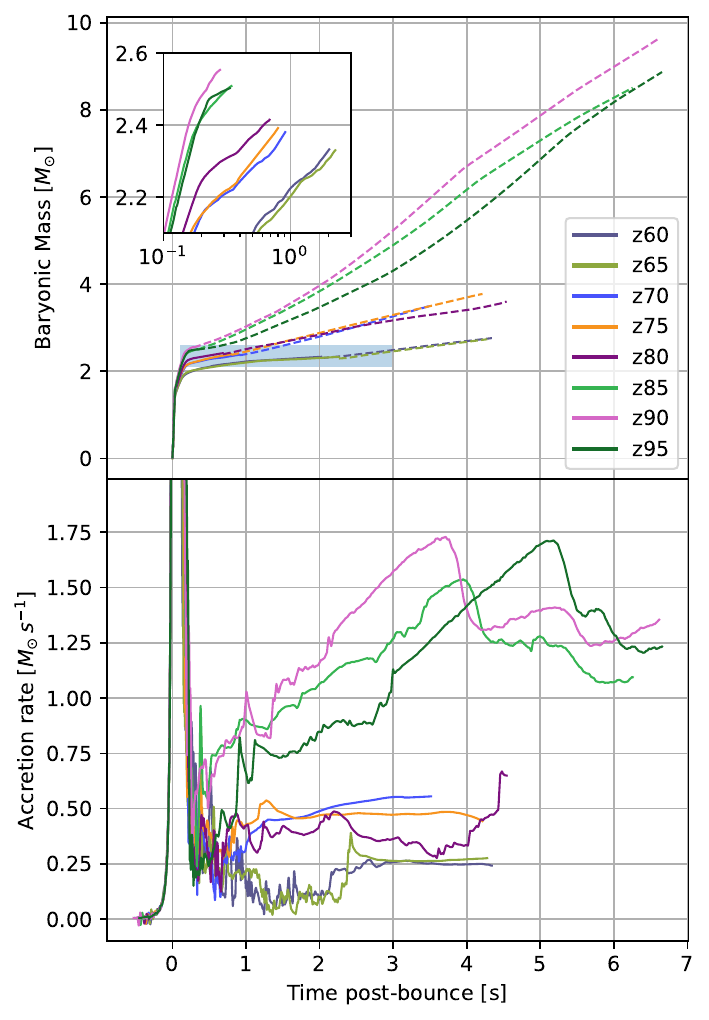}
    \caption{Top panel: Baryonic masses of the PNS (solid) and black hole (dashed) as a function of time. The inset panel corresponds to the shaded region in the top figure showing only the PNS masses until the moment of black hole formation over a narrower mass range. Bottom panel: Mass accretion rate at a radius of $100\, \mathrm{km}$. The accretion rates have been smoothed using a Savitzky–Golay filter to improve the clarity of the figure.}
    \label{fig:ns_bh_mass}
\end{figure}

The evolution of the baryonic masses of the PNS and subsequent black holes are shown in Figure~\ref{fig:ns_bh_mass} by solid and dashed lines, respectively. PNS masses are calculated by integrating the mass over the volume where the density is greater than $10^{11} \, \mathrm{\mathrm{g} \, \mathrm{cm}^{-3}}$. The mass of the black hole cannot be calculated directly as a volume integral, and we instead determine the total baryonic mass accreted by calculating the mass on the grid outside the excision boundary. Accounting for the mass flux over the outer boundary, which is small on these timescales, any remaining change in mass on the grid is assumed to contribute to the mass of the black hole. 

The gravitational or approximate ADM mass $M_\mathrm{grav}$
is formally a specialisation of Equation~(\ref{eqn:full_adm}) to a conformal metric with the apparent horizon acting as the integration surface instead of spatial infinity, cp.\ \citet{Sykes:2023}. It takes the form,
\begin{equation}
\label{eqn:mgrav}
   M_\mathrm{grav} = -2R_\mathrm{BH}^{2} 
    \left.\frac{\pd \phi}{\pd r}\right|_{R_\mathrm{BH}},
\end{equation}
and is thus sensitive to the numerical solution of $\phi$, and in particular the use of a scheme that conserves the ADM mass of the entire domain~\citep{Sykes:2023}. Note that as long as there is matter around the black hole, formula (\ref{eqn:mgrav}) only provides an estimate for the final black hole mass in vacuum.

For black hole formation cases, the gravitational mass is ultimately the more relevant quantity as it is a measurable property of the black hole, whereas the baryonic mass swallowed by the black hole is not. However, up to small integration errors, the baryonic mass is continuous during the collapse of the PNS to a black hole, which is not guaranteed numerically for the gravitational mass.

Both baryonic and gravitational masses are presented in Table \ref{tab:coconut_results_table}. The PNS masses are the final masses taken the moment before they collapse to a black hole. The PNS surface satisfying $\rho = 10^{11} \, \mathrm{g \ cm^{-3}}$ is also used as the integration surface for its gravitational mass. The gravitational mass for the black hole is calculated similarly but using the excision surface as the integration surface.

PNS masses are, barring some stochastic variations, ordered by progenitor mass. More massive models produce a heavier PNS which collapses more quickly, while the PNSs made by less massive progenitors PNS survive for much longer -- over two seconds post-bounce for the z60 and z65 models -- and collapse at a smaller final mass. 
This is expected as these smaller PNSs have more time to cool and hence have weaker support by thermal pressure, whereas the rapidly accreting PNSs in the more massive progenitors are stabilised above the maximum mass for cold neutron stars by thermal effects. Both ADM masses (PNS and black hole) are lower than their corresponding baryonic masses due to the implicit assumption of a vacuum outside the integration surface not being upheld: the rest of the star is present, and this induces some additional curvature to the spacetime not accounted for in this calculation.

Accretion rates onto the remnant are highest in the more massive z85, z90 and z95 models, where the rate exceeds $1 \msun \,\mathrm{s}^{-1}$ for several seconds. The least massive progenitors, z60 and z65, have the lowest accretion rates. This tentative ordering of the accretion rates by ZAMS mass is the result of less massive models injecting energy into the gain region for a longer period of time due to the longer-living PNS. This drives more material outwards and slows down the accretion.

\renewcommand{\floatpagefraction}{.8}%

\begin{figure}
    \centering
    \includegraphics[width=0.92\linewidth]{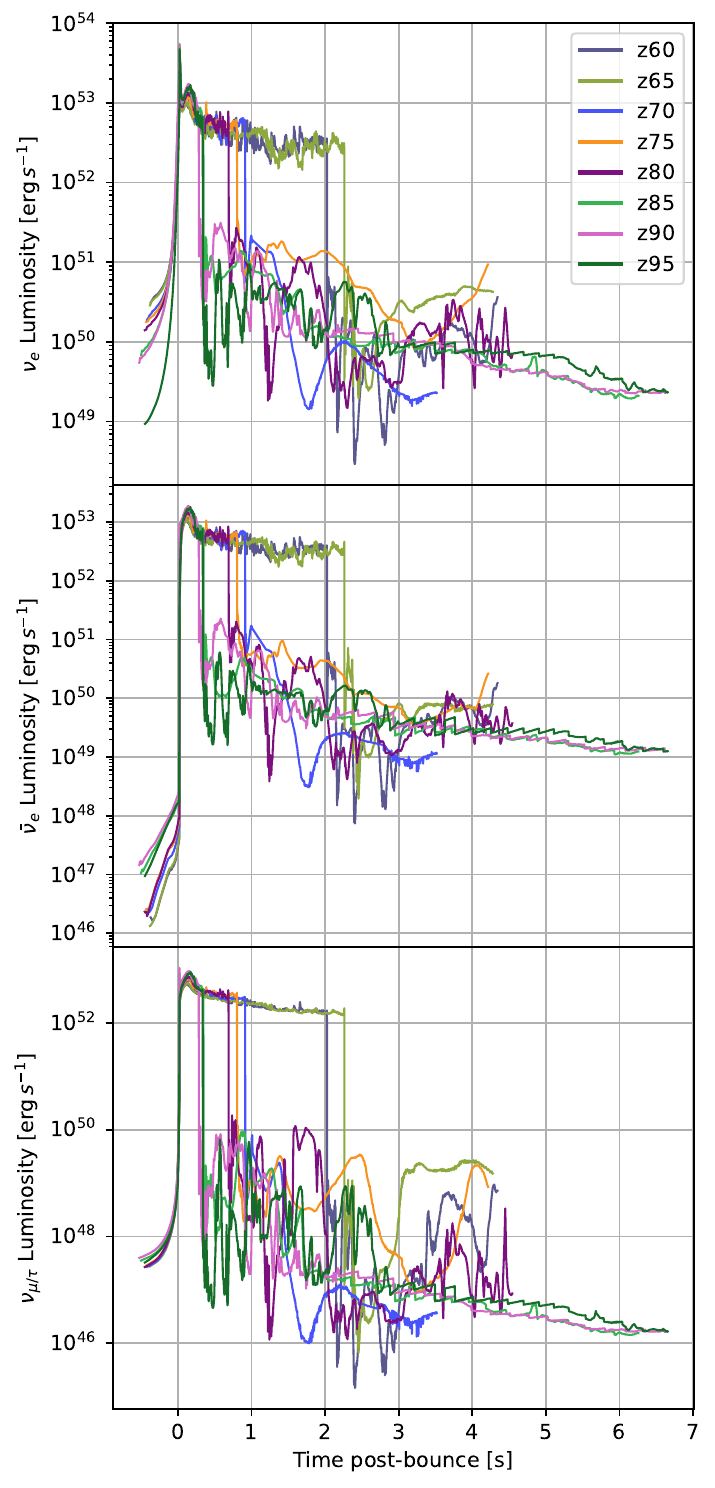}
    \caption{Luminosity of neutrinos from the onset of collapse to the end of the simulations. The top, middle and bottom panels show electron neutrinos, electron antineutrinos, and heavy flavours ($\mu$ and $\tau$), respectively. Time is normalised to bounce. 
    }
    \label{fig:neutrino_lum}
\end{figure}

\begin{figure}
    \centering
    \includegraphics[width=0.92\linewidth]{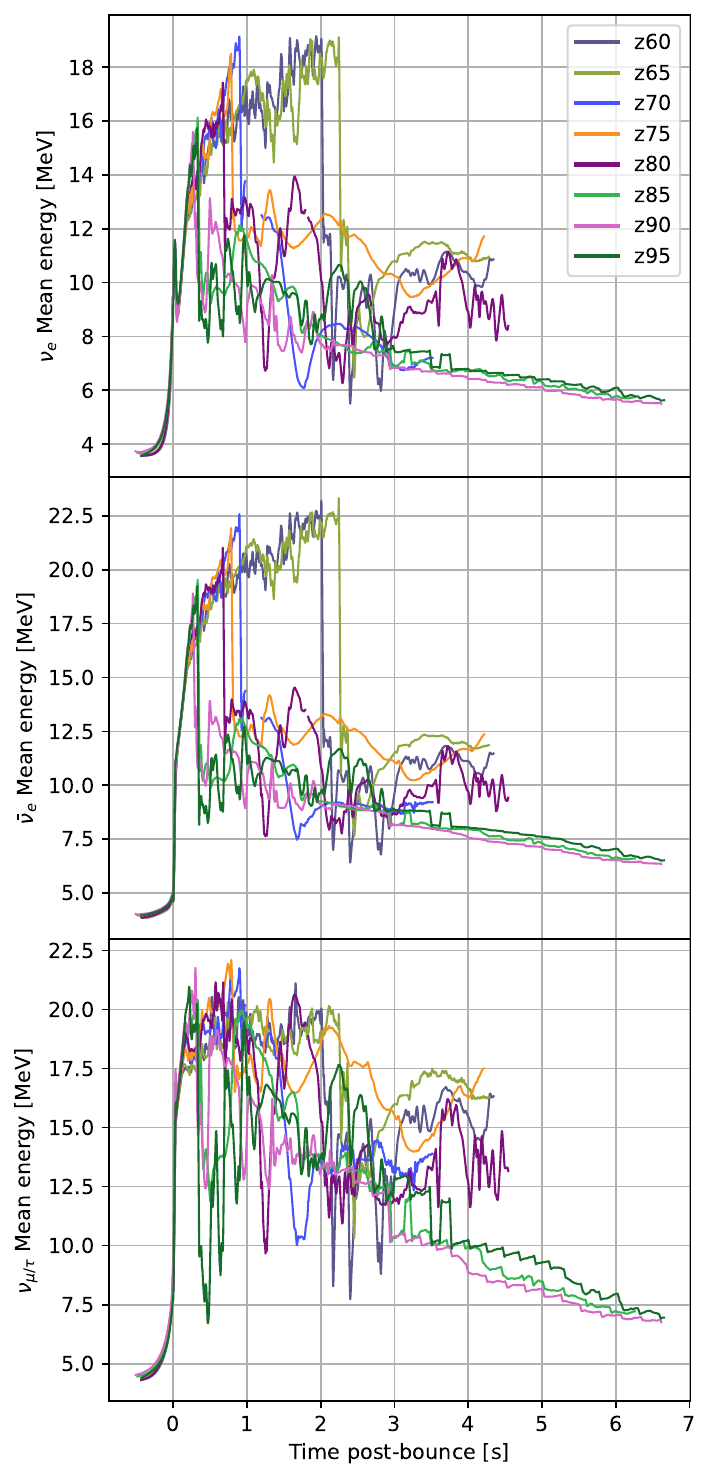}
    \caption{Neutrino mean energies for all species at the outer grid boundary. A Savitzky–Golay smoothing filter is applied to remove high frequency oscillations. Time is measured relative to bounce. The sawtooth pattern at late times, e.g., in the z95 model is due to the apparent horizon repeatedly
    moving by one zone, which slightly affects the solution of the neutrino transport in the vicinity of the black hole.
    }
    \label{fig:neutrino_mean_energy}
\end{figure}

\subsection{Neutrino emission}
The neutrino signature of black hole forming CCSNe is potentially a powerful probe of the core dynamics which cannot be studied via electromagnetic observations. The neutrino signal may be even more critical for failed CCSNe, where the only signal of the collapse may be neutrinos and gravitational waves from the core. Even if there is a visible explosion, as is the case in our simulations, the nature of the compact remnant -- neutron star or black hole -- can remain  unclear observationally; for instance it took over 30 years from the detection of SN1987A for the remnant to the conclusively identified as a neutron star \citep{Fransson_et_al:2024}. Neutrino detections provide a direct view of the fate of the PNS without relying on convoluted or indirect electromagnetic observation schemes.
To this end, we briefly comment on the neutrino signal here. 

The evolution of the electron neutrino, electron antineutrinos and heavy-flavour luminosities are shown in Figure~\ref{fig:neutrino_lum}. The electron neutrino luminosities peak at very similar values in the neutronisation burst at the models' respective times of bounce, $5-6 \times 10^{53} \, \mathrm{erg}\,\mathrm{s}^{-1}$. This is followed by a drop of the luminosity by about a factor of five on time scales of milliseconds to tens of milliseconds, and then a more gradual decline over hundreds of milliseconds to seconds, bringing the total decrease in the neutrino luminosity to about an order of magnitude below the peak value. If black hole formation happens sooner after bounce, the total decay in luminosity is less
as the collapse occurs at a higher mass accretion rate and accretion luminosity.

Heavy-flavour neutrino luminosity also decreases after bounce due to the contraction of the PNS and the corresponding shrinking of the neutrinosphere and increased redshift. Similar behaviour is also seen in the electron antineutrinos.

In all cases, black hole formation leads to a sharp drop in the neutrino signal. This is expected, as the neutrinosphere is quickly consumed by the black hole after it forms, and only volume emission from optically thin, lower-density matter continues. Previous work in 1D has shown this behaviour also \citep{Sykes:2023}, but the decrease in 2D appears slightly less extreme in some cases (e.g., for model z75), ranging between $2-3$ orders of magnitude; this is a similar range to that found by \citet{Rahman_Janka_Stockinger_Woosley:2022}. 
Unlike in 1D, there are significant fluctuations in the neutrino luminosities (not just electron neutrino but also electron antineutrino and heavy flavours) after black hole formation. These fluctuations can be up to about two orders of magnitude.

\begin{figure*}
    \centering
    \includegraphics[width=\linewidth]{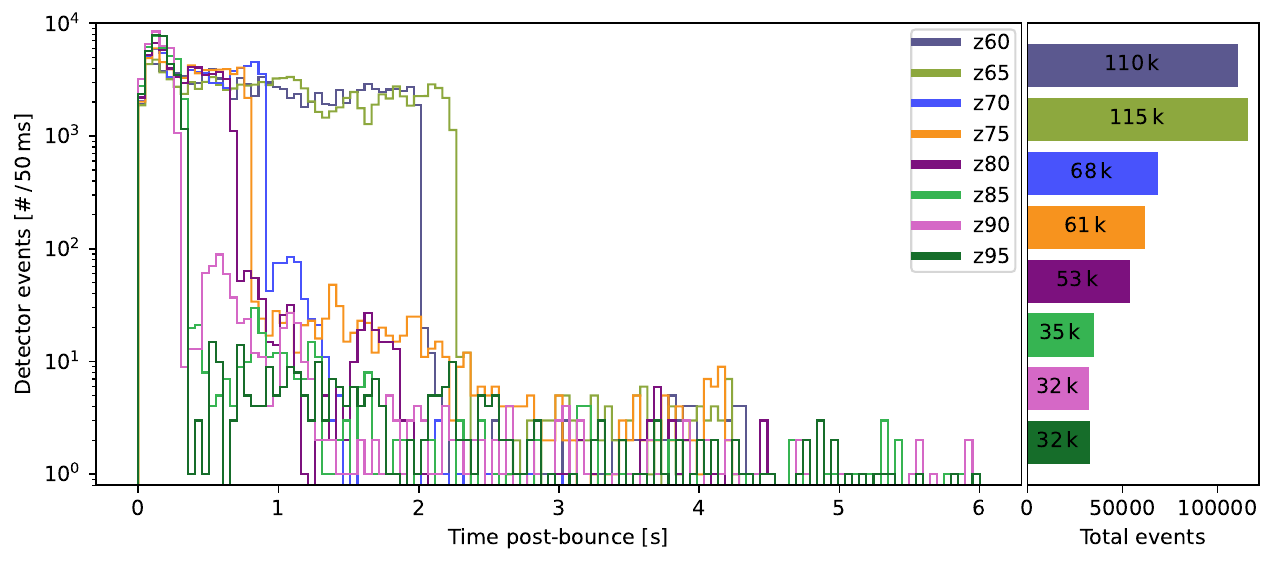}
    \caption{Predicted number of events per $50\,\mathrm{ms}$ window for each model for Hyper-K. Supernovae are set to a reference distance of $10\,\mathrm{kpc}$, neutrino oscillations are ignored, and the detector channels are limited to IBD only. The total number of neutrino events predicted for each CCSN is shown in the right panel}
    \label{fig:hyperk_rates}
\end{figure*}

We identify the emission region of the accretion flow as the zones with a positive gradient of neutrino flux with radius, which occurs just outside the apparent horizon of the black hole. These are the zones where there is still significant net production of neutrinos. We find that the mean density in these zones is strongly correlated with the variations. Such a correlation is expected since the emissivity from both electron or positron captures and bremsstrahlung is density dependent. For bremsstrahlung, the emissivity  $\eta$ increases roughly as $\rho^2$ to lowest order, but the dependence of all the rates on temperature and chemical potentials, as well as the detailed variation of the cross sections, generally imply a non-linear dependence of the emissivity on density, i.e., steeper than $\eta \propto \rho$.

However, it is more difficult to correlate these fluctuations with the accretion rates in Figure \ref{fig:ns_bh_mass}. This is a result of the neutrino fluctuations originating from dense downflows near the black hole which are particularly anisotropic. These downflows have small angular extent, but a radial extent of tens of kilometres; these regions are where increased neutrino emission occurs. Meanwhile, the accretion rates are calculated by integrating over spheres of constant radius. Thus, the contribution of some small solid angle of a downflow is small compared to the entire surface integral. This makes the accretion rate a poor indicator for ``clumped'' emission. 
This remains the case even if the reference radius for evaluating the accretion rate is adjusted to any alternative position between the apparent horizon and a few hundred kilometres.

This close to the excision boundary we have previously found numerical heating to be a potential issue \citep{Sykes:2023} with the power to overstate the neutrino signal by at least an order of magnitude in 1D models. In our current 2D models, the entropy does increase somewhat spuriously inside in the neutrino emission region by a factor of $\sim 2$. However, the density is quite well behaved in this region, and since the emissivity of electron neutrinos scales with $
\rho T^6 \propto \rho^{3} s^{2}$
for density, $\rho$ and entropy $s$ (under the radiation-dominated conditions in the downflows), the resulting factor of $\sim 4$ in variation from numerical errors in entropy does not explain the orders of magnitude fluctuations in neutrino luminosity. The variations in density (and associated thermodynamic variables) thus produce a real fluctuating neutrino signal, with the numerical increase in entropy providing only a subdominant effect.

Figure~\ref{fig:neutrino_mean_energy} shows the mean energies of the different neutrino species over time. The moment of black hole formation is clearly visible as a drop in mean energy across all species. This sudden drop should be an additional robust observational signature of black hole formation for neutrino detectors in addition to the drop in luminosity. The heavy-flavour neutrinos $\nu_{\mu}$ and $\nu_{\tau}$ have the highest mean energies after the black hole forms. All species, but particularly these heavy flavours, also oscillate in mean energy quite substantially during the black hole accretion phase; in the more massive models the heavy-flavour mean energy experiences brief rises back up to $\sim 20 \, \mathrm{MeV}$ -- on par with just prior to the collapse of the PNS. 

\citet{Kuroda_Shibata:2023} report a burst of high-energy heavy neutrinos in the milliseconds after black hole formation. Such a feature is not present in our models, which show oscillations of the heavy-flavour energy only as high as the post-bounce levels and far below the peak $\sim 80 \, \mathrm{MeV}$ reported. This may be result of \textsc{CoCoNuT-FMT} implementing non-isoenergtic neutrino-nucleon scattering \citep{Mueller_Janka_Marek:2012}, which transfers energy from heavy neutrinos to the background medium \citep{Keil_Raffelt_Janka:2003}. The code used by \citet{Kuroda_Shibata:2023} assumes isoenergetic scattering, thereby potentially explaining the discrepancy.

To address the possibility that this observable difference arises from failed vs. successful shock revival (as their model undergoes direct collapse, whilst ours explode), we compare their results to a simulation with modified neutrino physics, discussed in more detail in Section \ref{sec:expl_criterion}. This model does not explode, however once again we find no significant increase in the neutrino mean energies after black hole formation. Instead the neutrino luminosity after black hole formation slowly declines from a moderately high value (about $2 \times 10^{51} \, \mathrm{erg \, s^{-1}}$ in the case of $\nu_{e}$) while the bulk of the high-entropy shocked material accretes. After this, the luminosity is low and steady, similar to what was observed in 1D models \citep{Sykes:2023}. This suggests that the fluctuations in our primary simulations are a result of successful shock revival and the subsequent accretion of shocked material. Our data suggest that a burst of high-energy heavy neutrinos is unlikely to be a viable observational signature of black hole formation.

\subsubsection{Detection prospects}

For computing the prospective neutrino signal from the collapse of high-mass progenitor stars
to black holes in future neutrino detectors, we consider the planned Hyper-Kamiokande detector (Hyper-K) \citep{HyperK:2018} and use the \textsc{sntools} \citep{Migenda_et_al:2021} Monte Carlo event generator to estimate the detector signal for our models at a reference distance of $10 \, \mathrm{kpc}$. Since \textsc{CoCoNuT-FMT} does not output information on the neutrino energy spectrum by default\footnote{Simulations are run with multi-group transport with $21$ energy bins, but these are aggregated to neutrino energies, number densities and fluxes in the output.} we use the analytic prescription of 
\citet{Keil_Raffelt_Janka:2003} to describe the shape of the neutrino energy spectrum,
\begin{equation}
    f_{\nu}(E) \propto E^{\alpha} e^{-(\alpha + 1) E / \bar{E}}.
\end{equation}
This relies on the mean energy, $\bar{E}$, and a pinching parameter, $\alpha$, for which we use values from the high-resolution simulations of \citet{Tamborra_Mueller_Hudepohl_Janka_Raffelt:2012} in the post-bounce phase, giving $\alpha = 2.65,$ $3.13$ and $2.42$ for electron neutrinos, antineutrinos, and heavy flavours respectively. 

Figure~\ref{fig:hyperk_rates} shows the estimated event rates at Hyper-K for each of our models. Among the various detector channels 
in Hyper-K, we report the rates for 
inverse beta decay (IBD) as  the dominant channel, which is sensitive to the flux
and mean energy of $\bar{\nu}_{e}$. The electron scattering channel is subdominant, with around $5\%$ the events of the IBD channel.

For all models, the detector rate mirrors the corresponding luminosity in Figure~\ref{fig:neutrino_lum}. While many thousands of neutrinos are detected in the post-bounce phase, after black hole formation the rate typically falls to a few per second, perhaps up to a few hundred per second for the first few seconds of accretion in more luminous models.

Although the current study uses Population~III metallicity progenitors, their predicted neutrino signals can be used as templates for
prospective black hole formation events of
stars of similar mass that have been observed within or near our galaxy \citep{Schneider_et_al:2018}, including candidates for extremely massive pair instability supernovae \citep{Schulze_et_al:2024}. The sharp cut-off of neutrino events at black hole formation would be a strong indicator of a failed CCSNe within out galaxy. Furthermore, for a sufficiently close CCSN, a sharp drop followed by a dim, fluctuating neutrino signal could be a useful observational signature of collapse occurring after shock revival as opposed to direct collapse (which is expected to exhibit less asymmetry and a flatter neutrino luminosity \citep{Sykes:2023}).

\subsection{Emergence of the Final Mass Cut}
\label{sec:prometheus}

\begin{figure}
    \centering
    \includegraphics[width=\linewidth]{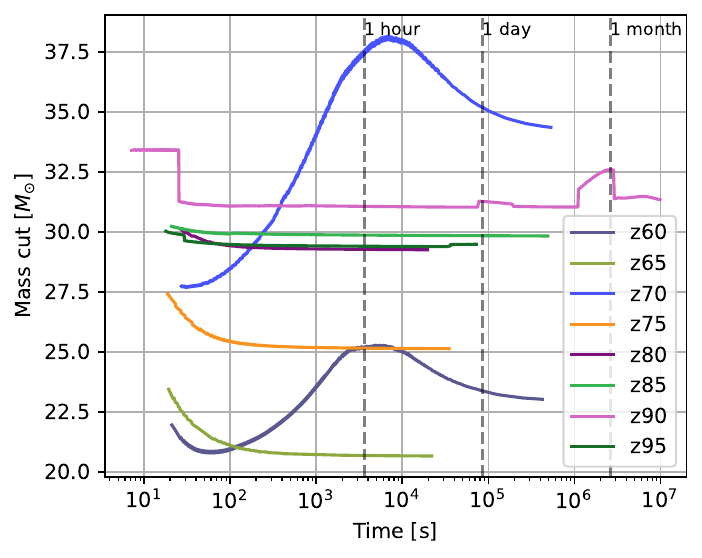}
    \caption{Evolution of the mass cut bifurcating the star into bound and marginally unbound ejecta. Times corresponding to one hour, one day, and one month are marked by dashed vertical lines for reference.}
    \label{fig:mass_cut_evol}
\end{figure}

Long-time simulations beyond shock breakout with \textsc{Prometheus} are run until the mass cut -- the point of innermost radius where $v_{r} > v_\mathrm{esc}$ -- is reasonably converged.
The evolution of this mass cut is shown in Figure~\ref{fig:mass_cut_evol}.
The formation of reverse shocks can make this difficult to determine, and some further fallback on significantly longer time scales
during the remnant evolution cannot be excluded.

Jumps in the nominal mass cut due to reverse shock dynamics are
particularly prominent in the z90 model at late times ($\sim 10^{6} \, \mathrm{s}$). In this case,
tracking the mass cut by the condition
$v_{r} > v_\mathrm{esc}$ may produce a discontinuous jump; this is because a reverse shock decelerates material as it travels, effectively rebinding it to the star and creating a pocket of slow material between the forward and reverse shocks. When the innermost edge of the reverse shock (i.e., the shock front itself) no longer satisfies the ejection condition, the new mass cut is then guaranteed to be somewhere ahead of this pocket of slow material, thus producing the upwards jump seen.

The next phase of this feature is a gradual increase in the nominal mass cut, as the shocked material transfers energy to the outermost layers of the star and is consequently slowed. This occurs for $O(10^{6}\, \mathrm{s}) $ -- up to about a month after the onset of explosion.

At this time, the mass cut suddenly jumps down again to a value only slightly larger than prior to the upwards jump. The reason for this downwards jump is a secondary shock originating from inner part of the star -- near the inner grid boundary. At some point the post-shock velocity of this secondary shock is sufficient to satisfy the ejection condition and the mass cut is abruptly moved deeper into the star. There are small variations after this, as the secondary shock moves outwards and the infall profile settles back to a stationary state.

The cause of this secondary shock is suspected to be a negative pressure gradient caused by the passage of the reverse shock which subsequently steepens into a forward shock wave, as described by \citet{Ertl_et_al:2016}. We do not investigate this phenomenon in detail however, and note that the reverse shock reaching the inner boundary causes a brief time period of subsonic flow over the inner boundary. This could potentially permit a numerical artefact in line with that initially suspected by \citet{Ugliano_et_al:2012}; however, the strong pressure gradient hints towards the effect being a real hydrodynamic feature of the explosion.

In models z60 and z70, the nominal mass cut
gradually moves outwards and then inwards in mass
in a manner distinct from the other models. This feature can be attributed to the structure of these particular progenitors, which have more compact and dense H/He envelopes, which
slow down the shock. In the other models, the mass in a radius of $\sim 10^{5} - 10^{8} \, \mathrm{km}$ is relatively small due to low densities in this region, whereas z60 and z70 have higher densities by a factor $\gtrsim 10$.

Consequently, when the shock moves through this region in the z60 and z70 simulations, it is decelerated by a greater amount, nominally rebinding some of the shocked material and increasing the mass cut. This is temporary however, as the shock continues to expand and transfer energy to the marginally bound outer shells, resulting in a deepening/decrease of the mass cut as additional material is swept up. More formally, this can be understood in terms of acceleration of the shock in response to variation of $\rho r^{3}$ in the dynamical structure of the star~\citep{Mueller:2020}.

Additional summary results from the \textsc{Prometheus} simulations are presented in Figures~\ref{fig:coco_prom_shock_r} and \ref{fig:exp_eng}. These show the evolution of the shock radius and explosion energy respectively for many days after the collapse.

\section{Criterion for successful supernovae}
\label{sec:expl_criterion}

In our set of simulations with standard settings in \textsc{CoCoNuT-FMT}, shock revival followed by black hole formation ultimately always results in an explosion.
However, this may not always be the case in nature, and long-time simulations to shock breakout are a relatively cumbersome approach to decide the final outcome of an explosion.
Ideally, one might therefore hope to decide by some analytic criterion whether a fallback supernova
with shock revival and subsequent black hole formation will ultimately explode.
It has been suggested by \citet{Powell_Mueller_Heger:2021} that the sonic point of the infall region defines the critical threshold for the shock on its path to a successful explosion. If the shock reaches the sonic point before the black hole forms, then it is likely that a weak shock travelling outwards at at least the speed of sound can outrun material accreting onto the black hole and eventually reach the stellar surface.
On the other hand, if the shock has not reached the sonic point, it is likely to be accreted with infalling matter, resulting in no observable explosion.

The formation of the black hole likely need not coincide exactly with the shock reaching the sonic point in the limiting case of a marginal explosion. The signal of the collapse of the core must propagate to the shock. 
Based on the sound-crossing time between the collapsing neutron star and the shock, one expects that the shock will take
$50-100 \, \mathrm{ms}$ to respond to black hole formation. This is a significant amount of time at this stage of the explosion, meaning that the shock can, in theory, be short of the sonic point by a few hundred kilometres and still make it into the
regime of subsonic infall pre-shock velocities before it is affected by the collapse to a black hole.

In our standard models, the shocks always reaches the sonic point well before the black hole forms. The closest case is the z90 model, where the black hole forms $40 \, \mathrm{ms}$ \textit{after} the shock reaches the sonic point. In our simulations $t_{\mathrm{BH}} / t_{\mathrm{sonic}}$ -- i.e. the ratio of black hole formation time and time at which the shock reaches the sonic point -- ranges from $1.2-9.4$. While this tests the positive case of the criterion -- that the explosion is successful after reaching the sonic point -- it does not test the converse, and nor does it identify the sonic point (give or take a sound crossing time) as the demarcation between successful and failed CCSNe. Therefore, these models are not suitable to conclusively probe the sonic point explosion criterion. However, with our excision code, we can test this criterion by artificially reducing neutrino energy deposition in the gain region.

For this purpose, we exploit the freedom to choose a closure relation for the variable Eddington factor, $p$, and the flux factor, $f$, defining the transition from diffuse to free-streaming limits in \textsc{CoCoNuT-FMT}. The previously implemented closure -- maximum entropy using the Fermi-Dirac distribution (MEFD) with maximal packing -- takes the form of a quadratic function, similar to other simple closures \citep{Murchikova:2017} such the Wilson \citep{LeBlanc_Wilson:1970} and Kershaw \citep{Kershaw:1976} closures. 

With this in mind, we define a general quadratic closure, $p(f)$, obeying both diffuse and free-streaming limits, namely $p(0) = 1/3$ and $p(1) = 1$ respectively. Additionally, a free parameter, $a$, allows us to vary the closure systematically as needed. While the conditions on closure relations of \citet{Anile_Pennisi_Sammartino:1991} require that $p'(1) = 2$ for correct signal propagation in the free-streaming limit (which produces $a = 4/3$, i.e., the MEFD closure with maximal packing), we disregard this condition in our experiment in preference of the flexibility afforded by varying $a$.

These conditions yield our general quadratic closure,
\begin{equation}
    \label{eqn:gen_closure}
    p(f) = af^{2} + \left(\frac{2}{3} - a\right)f + \frac{1}{3}.
\end{equation}

At high optical depths the two-stream Boltzmann closure of \citet{{Muller_Janka:2015}} is used. However once the flux factor increases above the singular point at $f = \sqrt{1 / (3a)}$ \citep{Muller_Janka:2015}, the code switches to the more common M1 closure method with Equation \eqref{eqn:gen_closure} as the closure.

\begin{figure}
    \centering
    \includegraphics[width=\columnwidth]{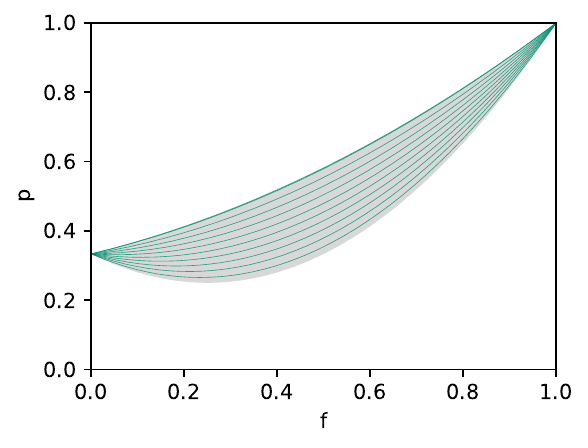}
    \caption{Closures used for exploring the conditions
    of continued shock expansion black hole formation
    in Section~\ref{sec:expl_criterion}. The grey region indicates the range $1/3 < a < 4/3$ while green lines show specific closures used to run simulations. }
    \label{fig:closures_fp}
\end{figure}

The free variable in the general quadratic closure is confined to a range, $1/3 < a < 4/3$. The resulting closure relations are shown in Figure~\ref{fig:closures_fp}. The lower limit guarantees a transition point to the M1 method exists in the $0 < f < 1$ bounds, which ensures that the general closure is accessible to regions of low optical depth. Meanwhile, the upper limit corresponds to the MEFD with maximal packing, which is a sensible lower boundary.

With this method, we are able to delay the onset of explosion such that it is closer to black hole formation. For most choices of the $a$ parameter, the explosion is still successful, however the $t_{\mathrm{BH}} / t_{\mathrm{sonic}}$ ratio is brought closer to one. One test simulation with the Kershaw closure resulted in almost coincident black hole formation and arrival of the shock at the sonic point; significantly, this model still exploded. The most relevant of our tests -- i.e. the one with the latest onset of explosion -- is the one with the extreme choice $a=0.341$. For this choice of $a$, the two-stream approximation of \citet{Muller_Janka:2015} is effectively used everywhere, and the region where the algebraic closure is used is squeezed out to very high flux factor. A consequence of this is a known shortcoming of the two-stream approximation -- that neutrinos become artificially forward beamed close to the neutrinosphere. Although this choice
of $a$ clearly gives a less accurate solution of the neutrino transport equation, it serves the purpose of reducing the neutrino heating enough to significantly delay shock revival.

In this model, the black hole forms $226 \, \mathrm{ms}$ after bounce, $40 \, \mathrm{ms}$ earlier than in our standard $z90$ simulation. Before black hole formation, the shock transiently expands, but does not reach the sonic point.
This is followed by complete fallback of the shock, with all of the high-entropy shocked material being accreted within $400 \, \mathrm{ms}$ of bounce. The time between black hole formation and transient shock expansion is $\sim 10 \, \mathrm{ms}$, at which point heating in the gain region is cut off. The diagnostic explosion energy is non-zero for about $170 \, \mathrm{ms}$, reaching $2.7 \times 10^{50} \, \mathrm{erg}$ at peak before vanishing.

Just prior to collapse, the explosion has developed protrusions of hot material along the central axis in both directions, as well as a ring-like structure around the equator. This structure collapses first, with the axial jets each following $40 - 50 \, \mathrm{ms}$ later.

Long-time follow-up simulations with the \textsc{Prometheus} code restore relic pair-instability pulses to the domain. The strength of these pulses, which steepen into weak shocks of energy up to $2 \times 10^{50} \, \mathrm{erg}$, may be influenced by gravitational mass loss due to neutrino emission \citep{Lovegrove_Woosley:2013, Ivanov_Fernandez:2021} during the collapse despite the pulses themselves being unrelated to the central explosion engine. Similar pulses, potentially in concert with binary interactions, are likely to have partially stripped the envelope at some point of the stellar evolution and highlight the need for more sophisticated stellar evolution models.

With these simulations, we have thus managed to probe the parameter space both for successful CCSNe with shocks that reach the sonic point, as well as for failed CCSNe with shocks that eventually turn around and are accreted by the black hole. However, the parameter space in the marginal region somewhere between the Kershaw closure's $a = 2/3$ and the edge-case $a = 0.341$ is clearly still undersampled in our study. Our results are consistent with the sonic point as the critical threshold for successful CCSNe, but more simulations are desirable to confirm this.

\section{Analytic approximations for shock evolution}
\label{sec:analytic_shock}

While simulations beyond black hole formation with excision are suitable for rigorously determining whether the shock survives during the early phase after black hole formation, full GR simulations until shock breakout are not numerically feasible and would not be cost-effective. The approach of mapping GR simulations to a Newtonian code and replacing the black hole with an accreting point mass is a relatively practical solution and indispensable for accurately modelling the interplay of mixing and fallback until and beyond shock breakout. However, this approach can still be cumbersome. Hence it would be desirable to extrapolate the eventual properties of the fallback explosion from the early phase after black hole formation by analytic theory.
Specifically, analytic estimates for the final mass cut of the ejecta, the explosion energy, and the black hole and the ejecta composition are of particular interest for relating to supernova transient observations, chemogalactic evolution, and observed black hole masses.

To test the possibility of capturing the long-time dynamics of fallback explosions in our simulations analytically, we follow \citet{Matzner_Ro:2021} and consider an approximate invariant for the propagation of \textit{weak} shocks\footnote{Strictly speaking, $\Delta v \ll c_\mathrm{s}$ is required for the velocity jump $\Delta v$ in the shock for the theory of weak shocks to be applicable.},
\begin{equation}
    R_{\pm} = \sqrt{\rho_{0} c_{0} A} \, \Delta J_{\pm},
    \label{eqn:rplus}
\end{equation}
where $J_{\pm} = v \pm 2c / (\gamma - 1)$ are the Riemann invariants along the outgoing and ingoing characteristics. The `0' subscripts indicate pre-shock quantities while `$\Delta$' indicates the change of quantity across the shock. On the timescales of our \textsc{CoCoNuT-FMT} simulations, there is very little change in the structure of the star outside $\sim 10^{4} \, \textnormal{km},$ and so the `0' subscript, pre-shock quantities correspond quite well to the initial value in the progenitor data. Assuming negligible reflection of the shock at shell interfaces\footnote{This typically holds as long as the shock is strong ($\Delta v > c_{s}$) by the time it reaches the H/He shell interface, where reflections tend to be stronger \citep{Mueller:2020}}, $\Delta J_{-} = 0$ from which it follows that $\Delta J_{+} = 2 \Delta v$. $R_{+}$ is not strictly conserved in practice \citep{Matzner_Ro:2021}. We investigate to what extent the approximation of constant $R+$ holds and possible solutions to account for changes
in $R_{+}$ in Section~\ref{subsec:shock_invariant}. 

The theory of \citet{Matzner_Ro:2021} suggests that the mass cut can be extrapolated from the state at the end of the
\textsc{CoCoNuT-FMT} simulations as follows:
\begin{enumerate}
    \item Calculate $R_+$ from the final state of the shock in the \textsc{CoCoNuT-FMT} simulation; at this point it is usually appropriate to assume spherical symmetry.
    \item Based on the density and sound speed of the progenitor, identify the radius at which the shock becomes strong again; i.e., $\Delta v \approx c_{s}$.
    Equation~\eqref{eqn:rplus} can immediately be used to do this.
    \item The velocity of the shock at the crossover to the strong-shock regime is estimated following \citet{Matzner_Ro:2021},
    \begin{equation}
        v_{s} = \bigg(\gamma + 1 + \sqrt{17 + \gamma(2 + \gamma)} \bigg)\frac{c_{0}}{4}.
    \label{eqn:vs_strong}
    \end{equation}
    \item Use a scaling relation for self-similar strong shocks, viz., $v_{s} \propto \rho_{0}^{-\beta}$, where $\beta = 0.19$ \citep{Whitham:1958, Sakurai:1960},
    to extrapolate the shock velocity until the ejection condition $f_\mathrm{sph} C v_\mathrm{s} = v_\mathrm{esc}$ is met. In this condition, $C$ is the post-shock acceleration factor \citep{Ro_Matzner:2013} and $f_{\mathrm{sph}}$ is a geometric factor for converting the ejection condition from one for planar flows to one for spherical flows.
    \item The final mass cut is then simply the minimum radius (or mass coordinate) where the ejection condition has been met. Assuming negligible mixing, the composition of the ejecta can then be determined by assuming all material outside the mass cut is ejected.
\end{enumerate}

\begin{figure*}
    \centering
    \includegraphics[width=\textwidth]{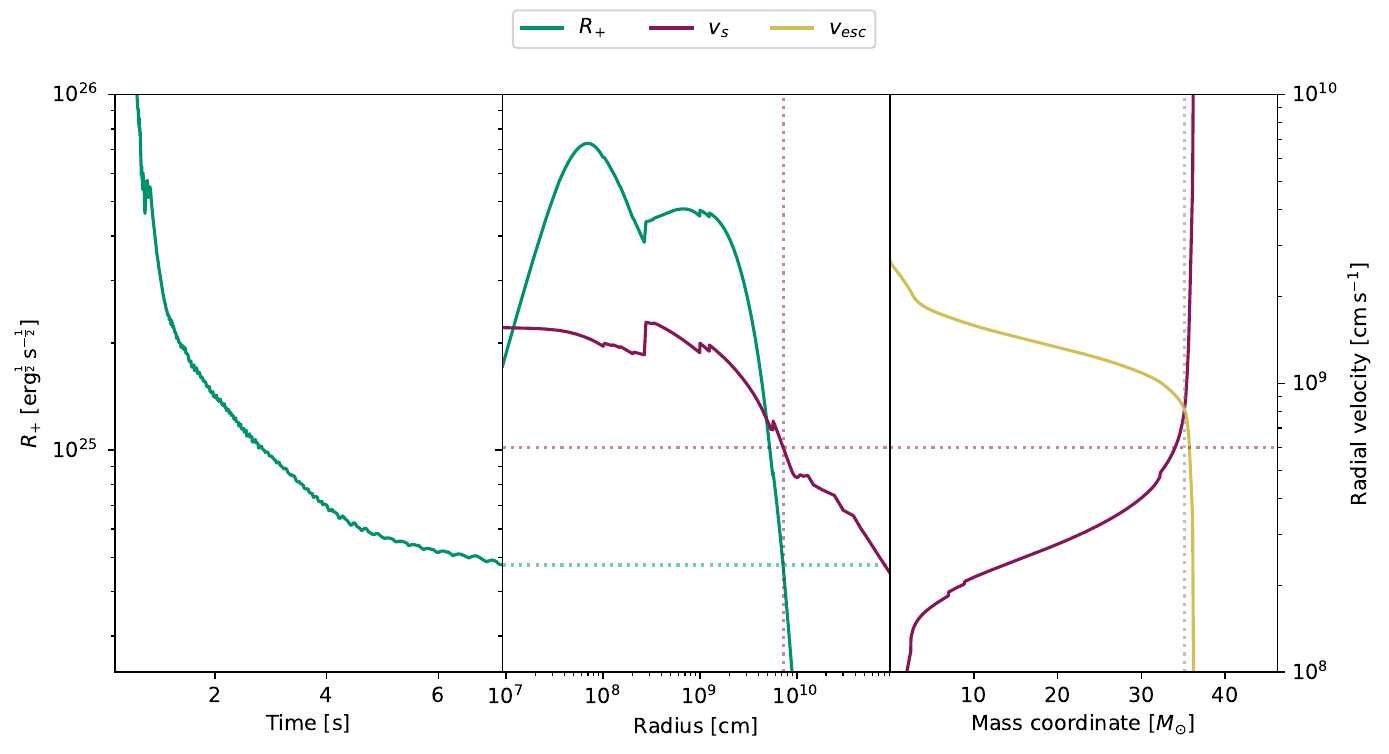}
    \caption{Illustration of the analytic method for extrapolating shock propagation for model z90. The left panel shows the time evolution of $R_{+}$ for the shock in the \textsc{CoCoNuT} simulation (scale on the left). 
    The central panel is used to determine when the shock leaves the weak shock regime again. To this end it shows $R_{+}$ for a velocity jump $\Delta v = c_\mathrm{s}$, which marks the boundary between the weak and strong shock regime. This critical curve for $R_{+}$ can be computed directly from the progenitor structure. 
    The purple line in the centre panel is the analytically estimated shock velocity $v_\mathrm{s}$ (scale on the right) for a shock with unit Mach number at different radii (Equation~\ref{eqn:vs_strong}). In the right panel, the purple line is the velocity of the shock according to the strong shock scaling law, $v_{s} \propto \rho_{0}^{-\beta}$. The yellow line shows the critical shock velocity for mass ejection
    $v_\mathrm{s,crit}= v_\mathrm{esc}/(f_\mathrm{sph} C)$.(see Section~\ref{sec:analytic_shock} for details). The dotted lines show how these quantities are connected. In the left panel, $R_{+}$ at the end of the simulation is the conserved value used in the weak regime; this is denoted by the dotted green line in the centre panel. The intersection of this dotted line with the green $R_{+}$ line in the centre panel tells us the radius at which the strong shock condition is reached, i.e., when the velocity jump at the shock becomes supersonic. A vertical purple dotted line is then used to show the estimated shock velocity (solid purple) at this radius. This shock velocity is carried over to the right panel through the horizontal purple dotted line and is used to determine the constant of proportionality for the strong shock scaling relation -- this is indicated by the intersection of this dotted line with the solid purple line in the left panel. The shock velocity can then be determined at arbitrary radii. The intersection point of this shock velocity curve and the escape velocity (including the correction terms) signals the point where the ejection condition is met. Thus, the final mass cut is represented by a grey vertical dotted line from the intersection point to the mass coordinate axis.}
    \label{fig:weak_method}
\end{figure*}

Figure~\ref{fig:weak_method} gives a visual representation of the extrapolation process for the z90 model.

\subsection{Evaluation of approximate shock invariant}
\label{subsec:shock_invariant}

The conditions for weak shock theory are not perfectly fulfilled in fallback explosions, and effects such as shock dissipation can become non-negligible.
Complications such as mild asymmetries may further limit the accuracy of the
theory of \citet{Matzner_Ro:2021}. It is thus important to analyse the validity of the analytic extrapolation and its inherent assumptions, like the constancy of $R_+$.

\begin{figure}
    \centering
    \includegraphics[width=\linewidth]{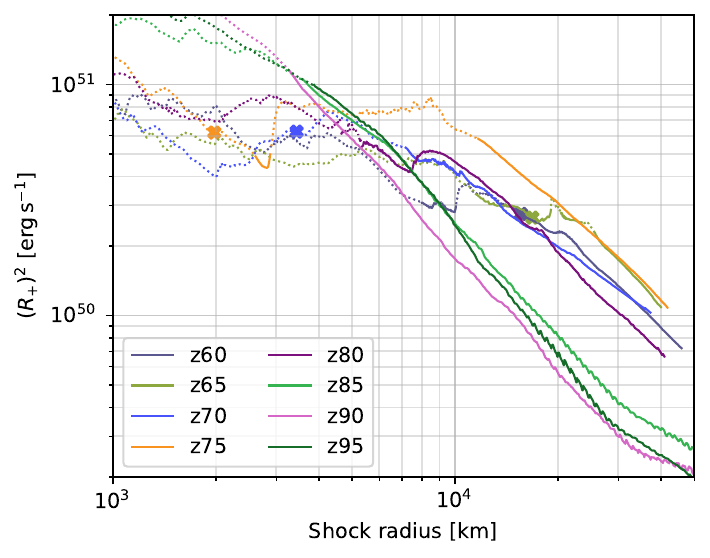}
    \caption{Value of $R_{+}$ as a function of shock radius. The time spent by the shock in the weak regime ($\Delta v < c_{s}$) is shown as a solid line (this is the region where the shock invariant theoretically holds), while the time as a `strong' shock is shown by dotted lines.}
    \label{fig:rp_radius}
\end{figure}

We therefore plot the squared value of the shock invariant as a function of mean shock radius in Figure~\ref{fig:rp_radius}. Evidently, $R_{+}$ is not well conserved in practice, which is expected due to dissipation. Nonetheless, if the point at which the shock transitions from weak to strong is not too far ahead of the shock radius at the end of the \textsc{CoCoNuT} simulation, then the deviation in $R_{+}$ may not be overly problematic; we will therefore proceed and investigate the consequences.

A prominent feature of Figure~\ref{fig:rp_radius} is that $R_+$ seems to more closely follow a power law with shock radius rather than being constant for a significant portion of the early shock evolution. This is particularly true for more massive models (e.g., z85, z90, and z95) which have a relatively constant slope for the entire duration of the weak shock phase until the end of the simulation. The less massive models have more variability in the beginning of the weak shock phase, either with a shallower slope (z70), or fluctuations in $R_{+}$ (z80). The z60 and z65 models are particularly interesting as the Mach number of the velocity jump across the shock is always close to unity for $r > 10^{4} \, \mathrm{km}$ (whereas the more massive models have Mach numbers less than $0.3$), resulting in a shock that is only barely classified as weak; in these cases power-law behaviour emerges quite late, at a large shock radius.

The times of black hole formation are marked by crosses in Figure \ref{fig:rp_radius}, although some form too early to be seen on the scale shown. The collapse of the PNS does not have a discernible effect on the conservation of $R_{+}$, i.e. by neutrino heating in the gain region which drives the shock.

The decay of $R_+$ leads us to instead investigate another empirically motivated invariant, $E_{s}$, of the form,
\begin{equation}
    E_\mathrm{s} = 2 \Delta v \sqrt{\rho_{0}r_\mathrm{s}^{3}}.    
\end{equation}
The motivation behind this is converting Equation \eqref{eqn:rplus} into an energy-like quantity. To achieve this, the $c_{0}$ dependence is removed and the area term converted to a volume term (constants such as $\pi$ are omitted for convenience). $E_\mathrm{s}^{2}$ then has units of energy.

\begin{figure}
    \centering
    \includegraphics[width=\linewidth]{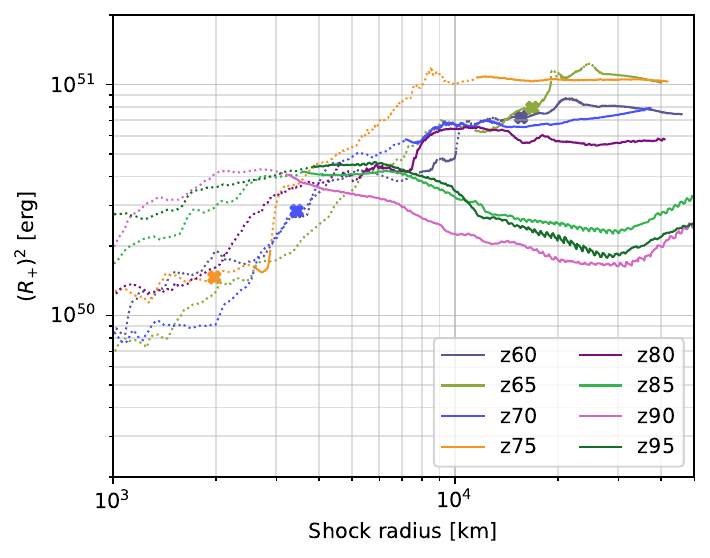}
    \caption{Same as Figure \ref{fig:rp_radius} but for the ``invariant'' $E_\mathrm{s}$.}
    \label{fig:es_radius}
\end{figure}

Figure~\ref{fig:es_radius} plots this alternative invariant as a function of shock radius for the duration of the weak shock phase of the \textsc{CoCoNuT} simulation. Most models show 
that this alternative invariant exhibits less variation than $R_+$ during the period of interest. The extrapolation of the mass cut based on $E_\mathrm{s}$  is completely analogous to the one for $R_+$. Although $E_\mathrm{s}$ is somewhat better conserved than  $R_+$, the following discussion mostly focuses on $R_+$, as this invariant has already been considered previously by \citep{Matzner_Ro:2021}. As we shall see, there is no clear advantage in using either invariant in terms of the accuracy of the predicted mass cut.

\subsection{Analytic prediction of shock propagation}
With our shock invariants determined, we are able to implement the recipe described previously to extrapolate the propagation of the shock through the weak shock phase, and into the strong shock phase until the final ejecta can be determined. 

In our models, the weak shock transitions to a strong shock at a radius less than $10^{5} \, \mathrm{km}$ in all cases; this is verified by the corresponding \textsc{Prometheus} simulations. In our case, running longer relativistic simulations  in \textsc{CoCoNuT} to avoid the weak phase entirely is not feasible as boundary effects at the outer grid boundary at $10^{5} \, \mathrm{km}$ start influencing the shock at these later stages.

The mass cuts determined by our analytic approximations are presented in Table~\ref{tab:analytic_results}. On average, analytically extrapolated mass cuts deviate from the \textsc{Prometheus} results by a similar relative error of about $10.3\%$  regardless of whether $R_{+}$ or $E_\mathrm{s}$ is used as the weak shock invariant.

\begin{table}
    \centering
    \begin{tabularx}{\columnwidth}{Y|YYY}
        \hline
        Model & 
        \makecell{$M_{\textsc{Prometheus}}$ \\ ($\msun$)} &
        \makecell{$M_{R_{+}}$ \\ ($\msun$)} &
        \makecell{$M_{E_\mathrm{s}}$ \\ ($\msun$)} \\
        \hline
        z60 & 23.0 & 20.5 & 21.5\\
        z65 & 20.7 & 20.4 & 21.9\\
        z70 & 34.4 & 25.3 & 27.2\\
        z75 & 25.1 & 25.6 & 26.9\\
        z80 & 29.3 & 28.3 & 29.8\\
        z85 & 29.8 & 33.7 & 34.0\\
        z90 & 31.3 & 35.2 & 35.5\\
        z95 & 29.5 & 33.4 & 33.4\\
    \end{tabularx}
    \caption{Mass cuts calculated directly from the \textsc{Prometheus} simulations and by analytic extrapolation.}
    \label{tab:analytic_results}
\end{table}

\begin{figure}
    \centering
    \includegraphics[width=\linewidth]{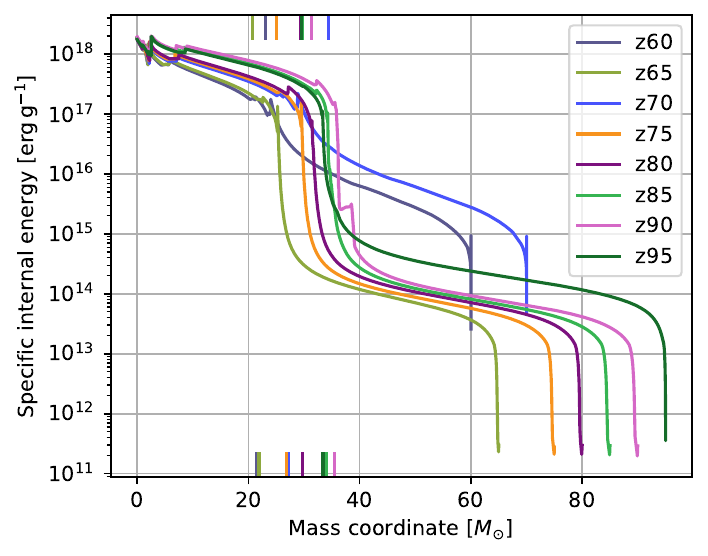}
    \caption{Specific internal energy as a function of mass coordinate from the \textsc{KEPLER} progenitor models. Coloured lines at the top of the figure show the mass 
    cuts from the \textsc{Prometheus} simulations for reference, while lines at the bottom show extrapolated mass obtained with the ``invariant'' $E_\mathrm{s}$.}
    \label{fig:kepler_int_energy}
\end{figure}

Figure~\ref{fig:kepler_int_energy} shows the specific internal energy profiles for each progenitor.
Aside from z60 and z70, all models show a sharp drop in internal energy typically somewhere in the mass coordinate range of $20\texttt{-} 40 \msun$ corresponding to the edge of the helium core.

Despite the approximate nature of the shock invariants, the analytic prediction of the mass cut turns out to be relatively robust and not too dissimilar to the simulation results. This can be understood from the structure of the progenitor stars.
The specific binding energy of the outer shells (i.e., the hydrogen envelope) is low, and hence the envelope can be easily ejected.
The simulated mass cuts (lines at the top of Figure \ref{fig:kepler_int_energy}) tend to fall a few solar masses inside this sharp drop, while the analytic mass cuts with the $E_\mathrm{s}$ invariant (bottom lines in Figure \ref{fig:kepler_int_energy}) are more tightly clustered around the drop itself -- ranging from slightly inside, to slightly outside the shell interface. The analytic estimates skew lower relative to the drop in specific internal energy for lower progenitor mass. This is because the less massive progenitors have more energetic explosions which can unbind material deeper inside the star instead of just the weakly bound envelope.

Within a margin of generally a few solar masses, the analytic mass cuts are a decent estimate of the actual mass cut in most models. Furthermore, analytic extrapolation seems robust even when the sub-optimal $R_{+}$ invariant is used, indicating that the mass cut is less sensitive to the shock dynamics in the weak shock regime. Instead it is the analytic treatment of the strong shock regime, and the progenitor structure itself, which has a significant impact on the mass cut with this method.

As the shock propagates in the outer layers of the He core, it encounters a strong density gradient (i.e., conditions akin to the surface layer of the star) and the assumptions of \citet{Sakurai:1960} for shock dynamics are naturally fulfilled. Thus it is unsurprising that the self-similar shock solution of \citet{Sakurai:1960} for strong shocks seems to perform well under the conditions we apply it.

The z70 model is the most prominent exception, with a $21\%-26\%$ deviation from the analytic prediction from the mass cut in the simulations. This discrepancy is likely the result of the more tightly bound envelope structure which causes the mass cut to be more sensitive to small changes in the shock and its extrapolation. The effect is less evident for the similarly structured z60 model; owing to the smaller size of this star, the final shock position for z60 in the \textsc{CoCoNuT-FMT} simulation is slightly further out in the star, meaning that it is slightly closer to the eventual mass cut to begin with. This means the model spends less time in the regime that requires analytic extrapolation, and consequently the errors in the extrapolation of the shock dynamics do not accumulate as much; hence the analytic prediction performs slightly better.

Interestingly, the smallest and largest mass cuts belong to the z65 and z70 models respectively, despite these being adjacent in ZAMS mass. This indicates that even with a homogeneous progenitor set such as that used in this work, the intrinsic scatter in remnant mass is large.

\begin{figure*}
    \centering
    \includegraphics[width=\textwidth]{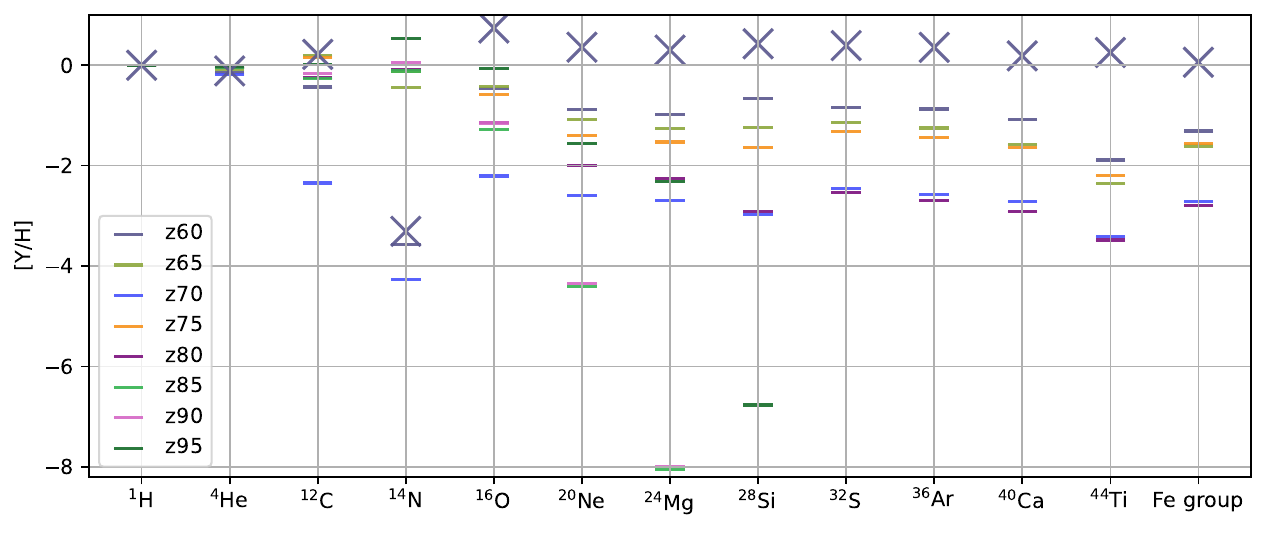}
    \caption{Composition of the  ejecta from \textsc{Prometheus} simulations expressed as abundances relative to hydrogen and normalised by solar values. An estimate of the ejecta composition of model z60 based on the analytically predicted mass cut is shown by grey crosses.}
    \label{fig:prom_ejecta}
\end{figure*}

\subsection{Ejecta composition}
Since CCSNe are thought to be the source of a significant fraction of the heavy elements in the Universe \citep[e.g.,][]{Woosley_Heger_Weaver:2002, Thielemann_Isern_Perego_vonBallmoos:2018, Kobayashi:2020}
, it is important to consider the composition of material produced by black hole forming explosions. The problem is particularly intriguing for Population~III stars, which are expected to show characteristically different yields \citep{Heger_Woosley:2010, Tominaga_Umeda_Nomoto:2007} that may explain peculiar abundance patterns in extremely metal-poor stars that were polluted just by a single or a few supernovae.

We show the composition of the ejecta from our \textsc{Prometheus} simulations in Figure~\ref{fig:prom_ejecta} using the usual bracket notation for element Z:
\begin{equation}
    [ \mathrm{Z}/\mathrm{H} ] = \log_{10} \bigg(\frac{Y(\mathrm{Z})}{Y(\mathrm{H})} \bigg) - \log_{10} \bigg( \frac{Y_{\odot}(\mathrm{Z})}{Y_{\odot}(\mathrm{H})} \bigg),
\end{equation}
where $Y = X/A$ is the composition fraction in $\mathrm{mol\,g}^{-1}$ with $X$ being the mass fraction and $A$ the mass number.
We use solar abundances
$Y_{\odot}$ from \citet{Lodders:2019}, appropriately integrated across isotopes.
We show abundances for hydrogen, $\alpha$-elements
up to ${}^{44}\mathrm{Ti}$ and iron-group nuclei collectively (since the total mass of iron-group ejecta is somewhat less sensitive to uncertainties related to the very simply treatment of nuclear burning and freeze-out from nuclear statistical equilibrium in our models than the detailed composition).

Models z60, z65, and z75 produce ejecta with the largest fractions of elements heavier than silicon. z70 and z80 also produce a notable amount of all tracked species, with the remaining models tending to eject higher proportions of lighter elements. 
The fact that some of the heavier elements are ejected at all indicates that a non-negligible amount of mixing occurs during the explosion of some models. Models with greater ejection of heavy elements tend to also have larger explosion energies.

\begin{figure}
    \centering
	\begin{subfigure}{\linewidth}
		\includegraphics[width=\linewidth]{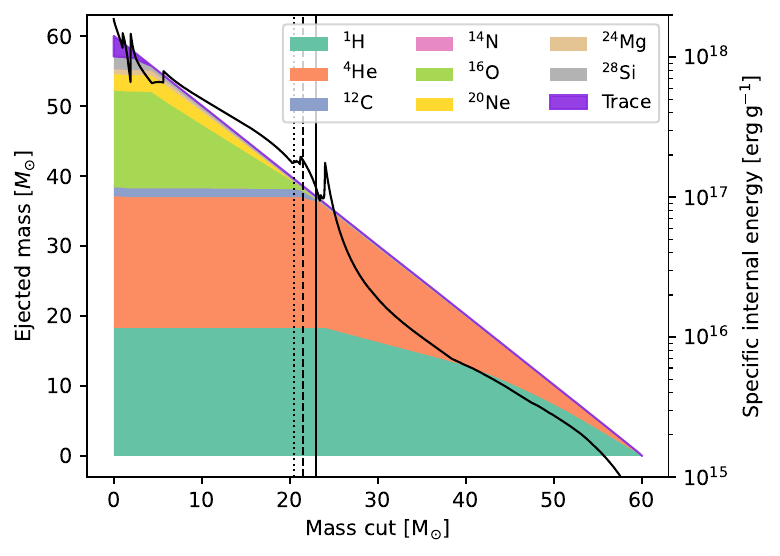}
		\caption{z60}
		\label{fig:z60_ejecta_mc}
	\end{subfigure}
	\begin{subfigure}{\linewidth}
		\includegraphics[width=\linewidth]{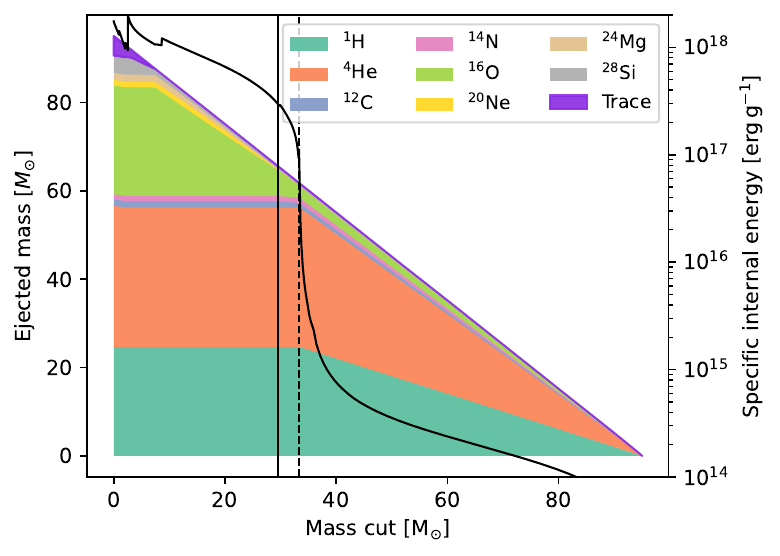}
		\caption{z95}
		\label{fig:z95_ejecta_mc}
	\end{subfigure}
    \caption{Estimated composition of the ejecta based on the progenitor structure for a given mass cut for the least and most massive progenitor models, z60 and z95. All tracked elements up to $^{28}\mathrm{Si}$ are shown, with all other species denoted by `trace', so called because there are only present in trace amounts outside the pre-collapse core. The internal energy of the progenitor is also shown (black curve and right axis). Three vertical lines show the mass cuts calculated in this work: Solid for mass cut from the \textsc{Prometheus} simulations, dotted for the extrapolation based on $R_\mathrm{+}$ method, and dashed for the extrapolation based on  $E_\mathrm{s}$.}
    \label{fig:ejecta_mc}
\end{figure}

Very small fractions of CNO elements in the z70 model are reflective of the large mass cut, which ejects primarily H and He. Meanwhile, the low $^{14}\mathrm{N}$ yield of z60 suggests a lack of H/He interface mixing.

For comparison, the ejecta composition may also be estimated using the analytically predicted mass cuts and the progenitor composition.
Figure~\ref{fig:ejecta_mc} visualises the predicted dependence of the ejecta
composition on the mass cut for models z60 and z95. These are the least and most massive progenitors in our study, and also display markedly different features in both the progenitor structure and explosion dynamics and timescales. The z60 (and also the z70) progenitor is up to a factor $10$ times more dense than other models above a radius of $\sim 2 \times 10^{5} \, \mathrm{km}$ with
a density profile that effectively flattens outside a mass coordinate of about $23 \msun$, with a sharp drop above $\sim 4 \times 10^{7} \, \mathrm{km}$. The other progenitors, in contrast, exhibit a more extended, power-law like profile out to large radii. Such extended envelopes with lower density can arise due to pair-instability pulsations.
In such cases, the outer H/He shells are also often only tenuously bound (see Figure~\ref{fig:kepler_int_energy}).

The mass cut in z60 mass lies just inside the lower boundary of the H/He envelope; only the two analytic ejecta estimates predict any sizeable ejection of C, N or O.
The position of the simulated or analytically predicted mass cut does not correlate strongly with any features in the internal energy profile, but seem to cluster near the outer edge of the metal core.

The analytic estimates of the mass cut for the z95 model coincide with a sharp drop in the specific internal energy at the inner boundary of the H envelope. The lack of scatter in the estimates again illustrates that in the case of these high-mass progenitors, uncertainties in the extrapolation of the shock
propagation for weak explosions play a minor role; regardless of uncertainties, the mass
cut will likely be placed at the edge of the He core.
Note that in both progenitors, the active or exhausted
He shell is quite thin, so the steep
density gradient effectively demarcates the
metal core from the H envelope.

We note that the composition corresponding to the solid vertical line in Figure~\ref{fig:ejecta_mc} does not exactly match the composition shown in Figure \ref{fig:prom_ejecta} since the former assumes no mixing during the explosion whereas the latter is obtained from the \textsc{Prometheus} simulation. The composition of the ejecta based on these analytic estimates is also shown for the z60 model in Figure~\ref{fig:prom_ejecta} for better comparison with the simulation results. The analytic estimates systematically overestimate the production of elements with mass number $A \geq 16$, and are also moderately inaccurate for lighter elements. Overall, they still permit zeroth-order estimates for the yields of heavy elements in the ejecta, but the accuracy of these estimates is limited because the analytic approximations do not exactly capture the shock dynamics and also do not account for mixing in the ejecta.

\section{Conclusions}
\label{sec:conclusion}

We have conducted long-time, general relativistic 2D simulations of eight massive black hole forming CCSNe for progenitors from $60 \msun$ to $95 \msun$. We have continued these models further to, and beyond, shock breakout by means of Newtonian simulations with a point mass to replace the black hole.

Our simulations have allowed us to probe the conditions for incipient supernova explosions to survive black hole formation. 
We find both shock revival and successful explosions for all of our progenitor models, and our long-duration simulations show significant mass ejection.
It has previously been proposed that the threshold for a successful explosion is met if the shock has reached the sonic point by the time of black hole formation \citep{Powell_Mueller_Heger:2021}
In all our models, the shock reaches the sonic point well before the black hole forms, and we therefore have insufficient data on failed supernova fully to assess this criterion. One simulation with modified neutrino physics which collapses after black hole formation offers tentative evidence that
explosions will fail if the shock has not reached
the sonic point by the time of black hole formation.
However, more work is needed across more of the progenitor parameter space to be sure.

Explosion energies range from $4.1 \times 10^{50} \, \mathrm{erg}$ on the low end for model z85, up to $2.5 \times 10^{51} \, \mathrm{erg}$ for the most energetic model z60. After black hole formation, explosion energies tend to reach fairly converged values after about $100 \, \mathrm{s}$. The black hole masses range from $2.4 \msun$ to $9.5 \msun$ 
by the end of the relativistic \textsc{CoCoNuT-FMT} simulations, and are
ordered roughly by increasing ZAMS mass at this point. After 
following the further evolution of the fallback explosion to and beyond shock breakout, the final black hole masses
range from $20.7 \msun$ to $34.4 \msun$. The mass cuts tend to cluster around the mass coordinate of the H/He shell interface, which are only loosely bound, especially in the high-mass progenitors 
that have previously undergone pair instability pulses.

We also investigated the neutrino emission as a potential multi-messenger signal from black hole forming supernovae. For all simulations, the pre-explosion phase and post-explosion PNS cooling phase is well above the noise threshold for events at a fiducial distance of $10 \, \mathrm{kpc}$. Black hole formation then results in a sharp drop in neutrino flux, the detection of which would be a clear signal of the formation of a black hole. Our simulations
suggest, however, that emission from the accretion flow
after black hole formation may still be detectable by next-generation neutrino observatories such as Hyper-K for 
sufficiently nearby event. In this phase, temporal variations in the accretion flow can cause low-luminosity fluctuations, which could provide information on the black hole accretion dynamics.

It is often desirable to extrapolate the outcomes of fallback supernovae from on the early explosion dynamics
before or shortly after black formation, without following the long-term evolution of the explosion in multi-D.
To this end, we investigated the possibility of determining the ejecta properties (mass and composition) with the help of analytic approximations for shock propagation in the weak and strong regimes.
We assessed two weak shock invariants $R_{+}$ \citep{Matzner_Ro:2021} and $E_\mathrm{s}$.
We found that, while $E_\mathrm{s}$ is better conserved during the weak shock regime, the performance of both invariants is comparable for the purpose of estimating mass cuts. The analytic extrapolation of the mass cut performs reasonably well, and is usually within about $10\%$
of the mass cut obtained directly from multi-D simulations.
This is due to the fact that the edge of the helium core
defines a natural structure feature where there shock
becomes strong enough to eject material. The structure
of the more massive progenitors, with very extended
and tenuously bound envelopes after pair instability
pulses, accentuates this behaviour. For progenitors
with less steep density gradients between the core and
the hydrogen envelope, the extrapolated mass cuts may
be less reliable.

The amount of ejected heavy elements varies considerably
across progenitors. While some of our models eject very little material aside from deeper layers of the star,
several of our Population~III progenitors eject elements from oxygen to the iron group in substantial abundances with $[Y/\mathrm{H}]\gtrsim -2$ and $[\mathrm{O}/\mathrm{H}]$
as high as 0 in model z95.

Our models represent a step towards a more systematic investigation of the explosion and remnant properties and multi-messenger signatures of fallback with end-to-end simulations than previous studies of individual multi-D fallback supernova models 
\citep{Chan_Mueller_Heger_Pakmor_Springel:2018,Chan_Mueller_Heger:2020}. Partly for computational convenience, the current study focused on rather massive progenitors and also utilized the LS220 equation of state that is conducive
to fast black hole formation, but is already marginally ruled out by observational constraints \citep{Oertel_Hempel_Klahn_Typel:2017, Fischer_et_al:2014, Lattimer:2019}. In future, it will desirable to scan a lower progenitor mass range that is representative of more likely fallback events expected in the Milky Way. This will require longer simulations until black hole formation occurs. Furthermore, the variety of electromagnetic transients that can result from fallback events need to be studied as well based on end-to-end simulations such as ours, in particular whether
they can account for extreme observed supernovae like
OGLE-2014-SN-073 \citep{Terreran_et_al:2017,moriya_18a}.
Due to the long time scales and physical complications that will eventually need to be included in \emph{three-dimensional} simulations -- such as rotation and magnetic fields -- a full understanding of fallback supernovae is, however, bound to remain a major challenge.

\section*{Acknowledgements}
The authors thank A.~Heger for helpful feedback.
This research is supported by an Australian Government Research Training Program (RTP) Scholarship. 
BM acknowledges support by
the Australian Research Council (ARC)
through grants FT160100035 and DP240101786.
We acknowledge computer time allocations from Astronomy Australia Limited's ASTAC scheme, the National Computational Merit Allocation Scheme (NCMAS), and
from an Australasian Leadership Computing Grant.
Some of this work was performed on the Gadi supercomputer with the assistance of resources and services from the National Computational Infrastructure (NCI), which is supported by the Australian Government, and through support by an Australasian Leadership Computing Grant.  Some of this work was performed on the OzSTAR national facility at Swinburne University of Technology.  OzSTAR is funded by Swinburne University of Technology and the National Collaborative Research Infrastructure Strategy (NCRIS). 
The colour scheme used in this work has been generated by \textsc{colourgorical} \citep{gramazio-2017-ccd}.

\section*{Data Availability}
The data from our simulations will be made available upon reasonable requests made to the authors.

\bibliography{bibliography}

\end{document}